# Modeling Water Transport Processes in Dialysis.

M.V. Voinova

**Abstract.**

Mathematical modeling is an important theoretical tool which provides researchers with quantification of the permeability of dialyzing systems in renal replacement therapy. In the paper we provide a short review of the most successful theoretical approaches and refer to the corresponding experimental methods studying these phenomena in both biological and synthetic filters in dialysis. Two levels of modeling of fluid and solute transport are considered in the review: thermodynamic and kinetic modeling of hemodialysis and peritoneal dialysis. A brief account for hindered diffusion across cake layers formed due to membrane filters fouling is given, too.

**Key words:** dialysis, transport, physico-mathematical modeling

Dedicated to Bengt Rippe

## Introduction.

In the current review we outline major efforts in providing quantitative description of modeling water transport processes in dialysis and present key mathematical models that address this issue, while also reviewing briefly the classical models for the membrane transport.

Biological filters of the kidney are notoriously complex. Artificial membranes that partially replace organism's renal function, being modeled after their natural prototypes, are also quite sophisticated. The present paper summarizes various theoretical approaches describing transport of fluid and solutes in artificial membranes, bridging physics and biophysics in order to collect different parts of the puzzle and bring light to the complex mechanism of membrane transport.

The review highlights key importance of mathematical modeling – a vital analytical tool which provides means for quantifying the dialysis procedure and thus increasing the quality of dialysis control. The review presents both hemodialysis and peritoneal dialysis modeling. The paper examines latest trends in mathematical modeling, which seek to solve the persistent problem of adequacy of dialysis prescription with computer simulations based on patients' statistic data, increased the accuracy of patient monitoring, and development of a new generation of device software.

The present review is aimed at both specialists (dialysis researchers, doctors working in the field of dialysis and blood purification, students of biomedical and life sciences) as well as wider audience with interest in nephrology.

## 1. Notes on the history of dialysis: the founding fathers.

### 1.1 Hemodialysis.

First hemodialysis (HD) treatment of human patient was conducted in Hessen in October 1924 by a German physician Georg Haas (*Dialisens historia2011, Paskalev 2001*). He used hirudin as the anticoagulant substance since heparin was not yet accessible (*Schmaldienst and Hörl 2004*). Haas's first dialyzer of U-shape collodion tubes was a construction immersed into a glass cylinder with bath solution (*Paskalev 2001*) with the first 15 minutes treatment time of the patient. In 1925 Haas wrote a short report describing the blood purification process performed on a human patient, the first hemodialysis attempt in the history of medicine (*Paskalev 2001*). Heparin became available in 1930s (*Schmaldienst and Hörl 2004*) and Haas used this anticoagulant in his experiments instead of hirudin. Collodion (or celloidin, cellulose-trinitrate) membrane method of preparation has been developed by Fritz Pregl, an Austrian chemist Nobel Prize winner in 1923 (*Pregl 1914*). Haas wrote:

"I have tried a series of different dialyzers from a variety of materials, animal and vegetable membranes and paper dialyzers. The best implementation was obtained from collodion with respect to fabrication, its dialysis effects, safe sterilization and because it can be obtained in any geometric shape."(from (*Paskalev 2001*)). Unfortunately, Haas's pioneering efforts in human dialysis were never acknowledged by the medical community and were consequently. It was another doctor, Willem Kolff of Kampen, the Netherlands, who made the next big step in the dialysis by constructing the dialysis machine, together with Henrik Berk.

The blood purification and dialysis procedure on animals was been developed at John Hopkins University, Baltimore, by the John Abel, Leonard Rowntree, and B. Turner. In 1913 the researchers published an article describing "vividiffusion" - removal of chemical sustances from the blood stream of animals in dialysis process (*Fresenius*). In their paper from 1914 (*Abel 1914*), the authors described a method when a blood taken outside a patient's body was purified and transmitted back to its natural vascular circulation system in a continuous flow (*Schmaldienst and Hörl 2004*) through a path isolated from the air (*Paskalev 2001)*:

"**...**Principle of the method is in connecting an artery of the animal by a cannula to an apparatus made from celloidin..in the form of tubes, immersed in a saline solution or serum and providing for the return of the blood to the animal's body by another cannula attached to a vein… The blood leaving the artery flows through a perfectly sealed system and returns to the body within a minute or two without having been exposed to contact with the air or any chance of microbial infection, while the diffusible substances which it contains can pass out, more or less rapidly through the walls of the tubes. Coagulation of the blood is prevented by injection of hirudin' (*Abel 1914)*.

Abel and colleagues named the device for dialysis the "artificial kidney'" (*Paskalev 2001, Schmaldienst and Hörl 2004*).

Willem Kolff, a physician from the Netherlands, used a rotating drum kidney constructed by Henrik Berk, an engineer, for treatment of a patient in 1945 (*Kolff and Berk*). The treatment took as long as one week and the result was successful: the patient (a 67-year-old woman with acute kidney failure diagnosis) lived for over 6 years after the procedure (and died after an unrelated illness). This successful attempt confirmed the effectiveness of therapeutical method based on the use of "artificial kidney" (*Haas 1952*) and a hemodialysis principle suggested by Haass and Abel's group. At that time cellophane was used as a new filtering material in the dialysis tubes filled with blood (*Fresenius*).

Another breakthrough in hemodialysis has been done by Swedish physiologist and inventor Nils Alwall of Lund University (*Dialysens historia 2011; Kurkus 2018; Kurkus etal 2007; Stegmayr 2016*). Nils Alwall's artificial kidney was a superior device when compared to the one designed by Kolff since it not only allowed to purify blood but also to remove the excess water after the dialysis procedure (ultrafiltration). The basic principle of ultrafiltration is to squeeze plasma water through filter membranes under pressure, after uremic toxins are removed during dialysis. The first patient (47-year-old man) was treated at the Lund hospital in September 1946. The 11-meter-long tubes that were used as filters for dialysis were made of cellophane – the same material used in the food industry at that time, e.g. for wrapping hotdogs.

Several technical problems appeared. For example, for the patients with a chronical kidney disease it was difficult to use cannules made of glass. The problem was solved in 1960, when doctor Belding Scribner together with surgeon Wayne Quinton of the U.S. created the arteriovenous shunt with two small teflon outlets leading to the artery and the vein (*Dialysens historia 2011*).

Another problem related to the first dialysis machine was the enormous space that the device was occupying. This has been improved by Frederik Kiil, a Norwegian physicist, who in 1960s managed to create a cellophane dialysis filtering system which was much more compact since only a small amount of blood taken from the body was used for the circulation in the dialysis machine (*Dialysens historia 2011*). Cellophane membranes, most common in 1970s, represented convenient and relatively inexpensive solution for dialysis as the filters could be reused in the process.

However, cellophane membranes were mechanically unstable and leaked during continuous dialysis process (*Schmaldienst and Hörl 2004*). Instead, in late 1960s, cuprophane and cellulosic-based membranes were proposed to be used as blood purification filters (*Schmaldienst and Hörl 2004*). In 1970s, the importance of biological tolerance of blood-contacting material was brought into focus **in** clinical tests of extracorporal devices (*Schmaldienst and Hörl 2004*). Among other physical properties, the hydrophobic-hydrophilic interactions of blood components with filter membrane play a key role. Since the original cuprophane membranes were found to be non-biocompatible, further research in this direction continued (*Hakim and Breillat 1984, Henderson 1983, Hakim and Fearn 1984*).

Cellulosic membranes are hydrophilic polymers. In water, due to the polar nature of cellulose molecules, a hydrogel layer is formed onto the membrane surface (*Schmaldienst and Hörl 2004*). Accordingly, it was found that the hydroxyl groups on the membrane surface could be partially involved in the complement activation (*Schmaldienst and Hörl 2004*). To minimize the activation of the complement cascades and to reduce the leukopenia, the second generation of the cellulosic-derived membranes was subsequently introduced for clinical use in the dialysis therapy. These improved materials were derived from cellulose substitutes such as cellulose acetate, cellulose diacetate and triacetate, as well as hemophane (*Schmaldienst and Hörl 2004*). Later efforts were directed towards the surface modification of membranes with the aim of improving their biocompatibility, with the complement system activation becoming the "golden standard" of the biocompatibility check (*Schmaldienst and Hörl 2004*). Simultaneously, in 1970s, the first synthetic membranes for hemodialysis using polyacrylonitrile (PAN)-based polymer (originally hydrophobic) was suggested by Rhône Poulenc (*Poulenc 2004*), after the experimental observation (*Babb 1975*) of desired enhanced permeability of membranes to higher molecular weight compounds in comparison with cuprophane filter retaining the to prevent amyloidosis (i.e., the long-term complication caused by serum $β_2$-microglobulin protein accumulation in a patient's blood at continuous HD).

Currently, both artificial polymer and cellulosic modified membranes are used in the dialysis therapy. We shall pay special attention to the modeling of water transport across artificial dialysis membranes in the sections 4 and 5 of present review.

**1.2 Peritoneal dialysis.**

Peritoneal dialysis is the alternative blood purification method applied in cases of severe chronical kidney disease (CKD), with peritoneum of the patient's abdomen used as a dialysis filter. When compared to hemodialysis, peritoneal dialysis is a less costly alternative. The first successful peritoneal dialysis was performed on animals by a German researcher, Georg Ganter in 1923. After this grand achievement, significant contributions to the method's development were made in the U.S., first by a Wisconsin trio of Wear, Sisk and Trinkle, who in 1936 suggested a system for a continual peritoneal dialysis, and then by Tenckhoff and Schenter, who developed a special bacteriologically safe abdominal catheter (*Tenckhoff 1968*).

In total, in the world at the moment the CKD therapy in numbers is approximately 1.6 million patients are treated with hemodialysis and about 200,000 – with peritoneal dialysis. In Sweden (according to 2009 statistics), 2760 patients were treated with hemodialysis, while 839 – with peritoneal dialysis (*Dialisens historia 2011*).

**1.3 The review structure**

Mathematical modeling of transport of water and solutes in both types of dialysis based on general thermodynamic principles of nonequilibrium thermodynamics which open the current review is presented in the following section 2. The classical two pore model and the distributed model are analyzed in details, too.

In the section 3, transport of water across natural membranes is considered in view of peritoneal membrane transport. In particular, the mechanisms of glomerular and peritoneal transport are compared. In parallel, both three pore model (TPM) of peritoneal membrane transport and the extended TPM theory with applications to the automated peritoneal dialysis are presented.

Section 4 is devoted to the transport across synthetic membrane filters used in hemodialysis. Mathematical modeling of molecular transport across synthetic dialysis membranes and diffusion across the tortuous membrane pathways in fouling (cake) layers is reviewed in section 5. The last two sections of the review composed of short discussion of theoretical models and approaches which are not presented in the main text as well as the outlook and summary.

The purpose of the current review is to provide the generalized theoretical physics view to the quantitative analysis of molecular transport in such enormously complex systems as natural and synthetic membrane filters in dialysis. The review is written with the hope to provide a theoretical support and facilitate the navigation of the interested readers in this important area of renal physiology – modeling complex processes inviolved in dialysis membrane filtration.

**2. Fundamental thermodynamic relations for the transport of fluid and solutes in dialysis.**

The fundamental concepts used in the description of transport phenomena in various media are famous Onsager's reciprocal relations. Due to the importance of these theoretical physics formulations for applications, we overview these concepts in a special theoretical physics subsection below.

**2.1 Nonequilibrium thermodynamics of irreversible processes**

Lars Onsager formulated the basics of nonequilibrium thermodynamics (*Onsager 1937*) including the reciprocal relations for the kinetic coefficients in transport phenomena described within the generalized forces –fluxes linear equations. In the thermodynamics of irreversible processes approach, for $n$ fluxes and all driving forces in the system, respectively, a system of equation can be written as following (*Kotyk and Janáček 1975*):

$$J_i = L_{ij} X_j \qquad (2.1)$$

$$i = 1, \ldots n$$

Here $J_i$ is a flux, $X_j$ is a conjugated force, $L_{ij}$ is the appropriate kinetic coefficient.

Onsager's principle of the symmetry of the kinetic coefficients states:

$$L_{ij} = L_{ji} \qquad (2.2)$$

In general, the symmetry of the kinetic coefficients or Onsager's principle reflects the deep-lying internal symmetry (*Landau and Lifshitz V5, 2009*) in the relaxation of the system not far from the thermodynamic equilibrium. In terms of the thermodynamically conjugate variables one can write:

$$Y_i = -\frac{\partial S}{\partial y_i} = \beta_{ik} y_k \qquad (2.3)$$

$$y_i = -\frac{\partial S}{\partial Y_i} \qquad (2.4)$$

where the defined quantities correspond to the entropy of the system, $(y_1, \ldots, y_n)$, as a function of $n$ physical variables $y_1, y_2, \ldots, y_n$ describing the system (*Landau and Lifshitz, V5, 2009*). With respect to the time reversal transformations, the relations (2.2) are valid only if physical quantities in (2.3), (2.4) both change sign (for example, when these variable are proportional to the velocities of some macroscopic motion in the system). In the case when only one of the quantities $y_i$ change sign under time reversal while the other $y_k$ remains unchanged, the Onsager's principle of the symmetry of kinetic coefficients gives (*Landau and Lifshitz, V5, 2009*):

$$\beta_{ik} = -\beta_{ki}$$

In particular applications, the linear relations (2.1) together with (2.2) form a basis for the fluxes-forces analysis in transport processes. Transport of water and solutes across filter membranes and the thermodynamic approach known as Kedem and Katchalsky theory (*Kedem and Katchalsky 1958*) considered in the section 2.2, is a special case of general relations (2.1, 2.2) where the kinetic coefficients obey the Onsager's principle:

$$J_1 = L_{11} X_1 + L_{12} X_2 \qquad (2.5)$$

$$J_2 = L_{21} X_1 + L_{22} X_2 \qquad (2.6)$$

In the phenomenological theory, the relations between the total volume flow of transported fluid and the exchange flow are (*Kotyk and Janáček 1975*):

$$J_V = L_p \Delta P + L_{pD} \Delta \Pi \qquad (2.7)$$

$$J_D = L_D \Delta \Pi + L_{pD} \Delta P \qquad (2.8)$$

with the Onsager's reciprocal relation for the coefficients:

$$L_{pD} = L_{Dp} \qquad (2.9)$$

The physical meaning of the kinetic coefficients is introduced in the Kedem and Katchalsky theory of transport of water and solutes across filtering membranes (see the section 2.2 in the review).

For a coarse nonselective membrane, in the absence of the hydrostatic pressure difference, $\Delta P = 0$, the volume flow vanishes:

$$J_V = L_{pD} \Delta \Pi = 0 \qquad (2.10)$$

When the osmotic pressure is absent, $\Delta \Pi = 0$, the exchange flux $J_D = 0$ and the ultrafiltration of nonselective membrane caused by the hydrostatic pressure only is absent:

$$J_D = L_{pD} \Delta P = 0 \qquad (2.11)$$

The selectivity of the membrane is introduced via assumption of the semipermeable membrane. For the latter, the osmotic flow and ultrafiltration play role which expressed in the non-zero cross coefficient, $L_{pD}$. In ideal semipermeable membranes,

$$J_D = -J_V \qquad (2.12)$$

where both volume and exchange fluxes are due to the water flow only (*Kotyk and Janáček 1975*) characterized by a single kinetic coefficient, the hydraulic conductivity $L_p$.

For the description of transport of dissolved substances, it is convenient to introduce a flow of solvent rather than the exchange flow. For the dilute solutions, Kedem and Katchalsky theory suggested for the flow of solvent $J_s$

$$J_s = (J_V + J_D)C_s \quad (2.13)$$

where $C_s$ is the (small) volume concentration of the dissolved compound.

The other coefficient introduced in the theory, is the reflection coefficient $\sigma$ (also called the Staverman reflection coefficient), defined as following:

$$\sigma = -\frac{L_{pD}}{L_p} \quad (2.14)$$

A system of two phenomenological equations (2.15, 2.16) incorporating both kinetic coefficients (*Kotyk and Janáček 1975*):

$$J_V = L_p(\Delta P - \sigma RT\Delta C_s) \quad (2.15)$$

$$J_s = \chi RT\Delta C_s + C_s(1-\sigma)J_V \quad (2.16)$$

where

$$\chi = (L_D - L_p\sigma^2)C_s \quad (2.17)$$

Equations (2.15-2.17) provide the complete thermodynamic description of transport of water and a single uncharged solute through the filter membrane.

### 2.2 Thermodynamics of membrane transport.

The theoretical and experimental studies of transport across membranes is a vast area of research, with issues ranging from thermodynamics of water and solute flows to cutting edge computational and nanotechnological tools exploring water and ion transportation on micron-scale and even nano-scale.

The traditional models for water transport account for the osmotic effects formalism developed in the early works of Kedem and Katchalsky (*Kedem and Katchalsky, 1958, 1961, 1963*).

Kedem-Katchalsky classical model based on the irreversible thermodynamics approach (Onsager's relations) describes the flow of solvent, $J_V$ and solute, $J_S$. For the semipermeable membrane, the solvent flow is given by the equation:

$$J_V = L_p S(\Delta P - \sigma \Delta \pi) \quad (2.18)$$

Here $L_p S$ is the hydraulic conductance of membrane of the surface area $S$ ($L_p$ is the so-called hydraulic conductivity or water filtration coefficient), $\Delta P$ is the hydrostatic pressure, $\Delta \pi$ is the osmotic pressure difference across the membrane, $\sigma$ is the reflection coefficient.

For the flow of solute, $J_S$, the Kedem-Katchalsky model states (*Kedem and Katchalsky 1961*):

$$J_S = J_V(1-\sigma)\bar{C} + PS\Delta C \quad (2.19)$$

In this equation $\bar{C}$ is the mean intramembrane concentration of solute, $\Delta C$ is the gradient of solute concentration and $P$ is the solute permeability coefficient. In the early publications (*Garlick and Renkin 1970, Renkin 1964, Renkin and Garlick 1970*), the expression (2.19) has been employed for evaluation of the capillary permeability to macromolecules (but using the simplifying assumption $\sigma = 1$).

The general expressions (2.18, 2.19) have been applied for various biological filters, from the simple semipermeable membranes to the complex filtration systems such kidney filtration barrier (*Tencer 1998, Deen 1985*) or transvascular walls (*Öberg and Rippe 2013*). In the latter system, the solute gradient $\Delta C (= C_p - C_i)$ value is the difference in the solute concentration in plasma, $C_p$, and in interstitium, $C_i$, respectively (*Rippe and Haraldsson 1994*). The mean solute concentration is, in general, a function of the fluid flow $J_V$ and of the product of the membrane solute permeability, $P$, and the membrane surface area, $S$.

The solute permeability coefficient $P$ is determined as a ratio of the (transcapillary) diffusion coefficient $D_S$ over the diffusion distance, $\Delta x$, so $P = \frac{D_S}{\Delta x}$.

The integration of equation (2.19) gives the following expression for the clearance $K$ value, which is the solute flux $J_S$ divided by the plasma concentration $C_p$ of solute:

$$K \equiv J_S/C_p = J_V(1-\sigma)\frac{1-\left(\frac{C_i}{C_p}\right)e^{-Pe}}{1-e^{-Pe}} \qquad (2.20)$$

Here the Péclet number $Pe = J_V(1-\sigma)/PS$ is introduced as a ratio of the solvent flow and the capillary diffusion capacity.

From (2.20), the capillary diffusion capacity, can be calculated at any given concentration ratio $C_i/C_p$ and for a given solvent flow (for example, a lymph flow), $J_V$, and the reflection coefficient of the membrane, $\sigma$:

$$PS = J_V(1-\sigma)/ln\left[\frac{\frac{C_i}{C_p}\sigma}{\frac{C_i}{C_p}-(1-\sigma)}\right] \qquad (2.21)$$

When the interstitial space is large and the solute concentration in there is small, respectively, one can get from (2.20) the simplified expression for the solute clearance $K$

$$K = J_V \frac{1-\sigma}{1-e^{-Pe}} \qquad (2.22)$$

The general equation for the clearance (2.20) may be rewritten as a sum of two terms:

$$K = PS\frac{Pe}{e^{Pe}-1}\frac{C_p-C_i}{C_p} + J_V(1-\sigma) \qquad (2.23)$$

The first to the right 'diffusive'' term vanishes in case of large Peclet number so the second ('convective'') term is dominated in the process. That means that the 'diffusive'' transport is very small at high volume flow values and the clearance:

$$K \approx J_V(1-\sigma) \qquad (2.24)$$

For the cylindrical pores, by applying Poiseuille's law, one can calculate the tissue capillary hydraulic conductance:

$$8\eta\, L_p S = \frac{A_0}{\Delta x}(\alpha \cdot r^2) \qquad (2.25)$$

Here $A_0$ is the total (unrestricted) area of pores in the membrane wall, $\Delta x$ is the unit diffusion path length, $\eta$ is the viscosity of water and $r$ is the radius of a cylindrical pore.

Another important physical characteristic of membrane transport is the solute diffusion coefficient, $D_s$. By definition, the diffusion coefficient is introduced as following:

$$D_s = \frac{RT}{6\pi N \eta a_s} \quad (2.30)$$

In there, $R$ is the gas constant, $T$ is the temperature (in Kelvin degrees), $N$ is the Avogadro's number and $a_s$ is the radius of solute.

Formulae (2.18-2.30) represent main thermodynamic relations for the macroscopic description of the transport processes in membranes.

**2.3. Two-pore membrane model.**

One of the widely used theories of transvascular transport is the so-called two-pore model, which combines the general non-equilibrium thermodynamics and kinetics of transmembrane transport for the analysis of convection and diffusion processes involved in the filtration of plasma solutes and water permeability. The basic assumptions and results obtained within the two-pore model are briefly summarized in the following two sections. The two-pore model has been used to describe permeability and transport phenomena in blood capillaries and different organs (*Bark 2013, Miller 2011, Taylor 1984, Ibrahim 2012*), glomerular sieving (*Rippe Asgeirsson 2006, Tencer 1998, Deen 1985*) and, recently, for the analysis of fluxes in synthetic dialyzing membranes (*Axelsson 2012*).

**2.3.1 Classical two pore- model.**

Classical two-pore model considers transport of water and solutes across a (capillary) wall through two pathways, distinct in size, i.e. large and small pores with a fixed diameter (*Arturson 1971, Blouch 1985, Axelsson Öberg 2012, Rippe and Haraldsson 1994*). According to this model, one can separate the transport across the membrane wall into two different pathways representing two kinds of pores – of small and large radii, $r_s$ and $r_L$, respectively. In this relation, the partitioning of fluid fluxes among multiple pathways may be considered within the homoporous or heteroporous membrane models (*Rippe and Haraldsson 1994*). In the homoporous approaches (*Renkin 1979, 1985, 1986, Garlick and Renkin 1970, Renkin 1964, Renkin and Garlick 1970*), linear flux equation (2) with added convective term has been analyzed with the so-called slope method (Perl 1975, Bark 2013, *Renkin 1977*) and, alternatively, (*Taylor 1977*) with the cross-point methods. These theoretical models were applied to miscovascular selectivity (*Perl 1975, Renkin Joyner 1977, Renkin Watson 1977*) and the steady-state lymphatic protein flux**es** analysis (*Renkin 1979, 1985, 1986, Renkin and Joyner 1977*).

For the transport of fluid in the so-called heteroporous membrane, one can find the generalization of the equation (2.18) for $m$ different-in-size pores (*Rippe and Haraldsson 1994*):

$$J_V = \sum_{i=1}^{m} J_{v,i} = \sum_{j=1}^{m} \alpha_i L_p S (\Delta P - \sum_{j=1}^{n} \sigma_{i,j} \Delta \pi_j) \quad (2.31)$$

Next step in the development of porous membrane models is the so-called distributed two-pore model.

**2.3.2 Distributed two - pore model.**

In distributed two-pore models it is assumed that the pore diameters are distributed with standard deviation (SD) around a mean diameter (*Öberg and Rippe 2014*). Since large pores are in most works considered as shunt non-selective pathways, only the small pores were taken into account (*Öberg and Rippe 2014*). However, recent experimental data in glomerular transport of large proteins (*references*) indicated the existence of upper size limit for molecules with radius about 110-115Å. This finding means that the large plasma proteins whose size exceeds the upper limit for pathways) for example, IgM) cannot normally penetrate across glomerular filtration barrier (GFB) (*Tencer 1998*).

Comparison between the classic and the distributed two-pore models shows that the log-normal distribution used in the classic (discrete) two-pore model shows a poor fit to the glomerular transport data in the region of radii between ~50-60Å (*Rippe Asgeirsson 2006*). Ficoll, a popular marker of GFB permeability, has been used in a number of studies (*Rippe Asgeirsson 2006*) for testing the two-pore distributed model. In most studies, the transport of solutes considered as a penetration of rigid spheres through the pore. Current experiments with Ficoll reveal that the flexibility of molecule plays a role in the transmembrane transport: more flexible Ficoll transport coefficients were different from, e.g. albumin, which behaves more like a rigid sphere (*Rippe Asgeirsson 2006; Venturoli Rippe 2005*).

Below the distributed two-pore model is presented in a brief (for the details of calculations, see (*Öberg and Rippe 2014*)).

Let us consider the steady-state transport (both diffusion and convection) of solute across a semipermeable membrane wall

$$J_{solute} = J_{diffusion} + J_{convection} = -DA\frac{dc}{dx} + J_V(1-\sigma)c \qquad (2.32)$$

In the equation (2.32), index $'s'$ stands for the solute,

In the equation (2.32), $D$ stands for the diffusion coefficient, and $A$ corresponds to the effective membrane area. The integration of the equation (2.32) gives us

$$J_{solute} = J_V(1-\sigma)\frac{C_P - C_i e^{-Pe}}{1 - e^{-Pe}} \qquad (2.33)$$

where the indices $'P''$ and $'i''$ denote plasma and filtrate concentrations, respectively. In there, the Péclet number (the 'convection –to - diffusion ratio')

$$P_e = \frac{J_v(1-\sigma)}{PS} \qquad (2.34)$$

The permeability-surface coefficient is defined as:

$$PS = \frac{D}{\Delta x}A = D\frac{A_o}{\Delta x}\frac{A}{A_o} \qquad (2.35)$$

The Stokes-Einstein equation gives the following expression for the diffusion coefficient:

$$D = \frac{kT}{6\pi\eta a_e} \qquad (2.36)$$

The sieving coefficient:

$$\theta = \frac{1-\sigma}{1-\sigma e^{-Pe}} \qquad (2.37)$$

Note, that solute clearance is the product of sieving coefficient $\theta$ and volume flux, $\theta \cdot J_V$.

To relate the ultrafiltration concentration to the plasma concentration, the authors (*Öberg, Rippe 2014*) use the following equation:

$$C_P = \theta \times C_i \qquad (2.38)$$

By using the log-normal distribution for the probability density function one can get:

$$g(r, u, s) = \frac{1}{r \ln(s) \sqrt{2\pi}} e^{-1/2 \left[\frac{\ln(r) - \ln(u)}{\ln(s)}\right]^2} \quad (2.39)$$

Then, the filtration barrier's net value of the the sieving coefficient $\theta$ is calculated as following:

$$\theta = f_S \frac{1 - \sigma_S}{1 - \sigma_S e^{-Pe_S}} + f_L \frac{1 - \sigma_L}{1 - \sigma_L e^{-Pe_L}} \quad (2.40)$$

where indices 'S' and 'L' correspond to small and large pores, respectively, $f_L$ denotes the fraction of liquid crossing the membrane via large pores, and the reflection coefficients

$$\sigma_S = \frac{\int_0^\infty r^4 g(r, u_S, s_S) \sigma_{h,S}(r) dr}{\int_0^\infty r^4 g(r, u_S, s_S) dr} \quad (2.41)$$

$$\sigma_L = \frac{\int_0^\infty r^4 g(r, u_L, s_L) \sigma_{h,L}(r) dr}{\int_0^\infty r^4 g(r, u_L, s_L) dr} \quad (2.42)$$

Analogously, corresponding Péclet numbers are given by the formulae:

$$Pe_S = \frac{J_{vS}(1 - \sigma_S)}{PS_S} \quad (2.43)$$

$$Pe_L = \frac{J_{vL}(1 - \sigma_L)}{PS_L} \quad (2.44)$$

Further, the permeability-surface coefficients can be calculated by using the following equations:

$$PS_S = D \frac{A_{O,S}}{A_O} \frac{A_O}{\Delta x} \frac{\int_0^\infty r^2 \left(\frac{A}{A_0}\right)_{h,S} g(r) dr}{\int_0^\infty r^2 g(r) dr} \quad (2.45)$$

$$PS_L = D \frac{A_{O,L}}{A_O} \frac{A_O}{\Delta x} \frac{\int_0^\infty r^2 \left(\frac{A}{A_0}\right)_{h,L} g(r) dr}{\int_0^\infty r^2 g(r) dr} \quad (2.46)$$

Now, one can find volume fluxes for each of membrane pathways (marked with index '$i$'). At the absence of osmotic pressure gradient:

$$J_{vi:\Delta\pi=0} = \alpha_i \times GFR \quad (2.47)$$

where $GFR$ denotes a glomerular filtration rate. When the osmotic pressure is non-zero, the Starling equilibrium equation for each of the pathway gives:

$$J_{vi} = \alpha_i K_f (\Delta P - \sigma_{i,net} \Delta \pi) \quad (2.48)$$

so the difference in volume fluxes is given by formula:

$$J_{vi,iso} = J_{vi} - J_{vi:\Delta\pi=0} = \alpha_i K_f (\sigma_{o,net} - \sigma_{i,net}) \Delta \pi \quad (2.49)$$

Here, the fractional hydraulic conductance values are introduced

$$\alpha_i = K_{f,i}/K_f \qquad (2.50)$$

and $K_f \equiv L_P S$ (Öberg, Rippe 2014).

By calculating the improper integrals in the denominators of equations (2.41- 2.44), authors found the analytical solutions (see, for the details of calculation, in (*Öberg and Rippe, 2014*) for the total cross-sectional area of porous GFB:

$$\frac{A_0}{\Delta x} = \frac{N}{\Delta x}\int_0^\infty \pi r^2 g(r)dr = \frac{N}{\Delta x}\pi G^2(u,s) = \frac{N\pi u^2}{\Delta x}e^{2\ln^2(s)} \qquad (2.51)$$

where N corresponds to the total number of pores per unit of the kidney GBF membrane.

By using Poiseuille law, one can calculate the hydraulic conductance which is given by the formula:

$$K_f = N\int_0^\infty \frac{\pi r^4}{8\eta \Delta x}g(r)dr = \frac{N\pi}{8\eta \Delta x}G^4(u,s) = \frac{N\pi u^4}{8\eta \Delta x}e^{8\ln^2(s)} \qquad (2.52)$$

For known $\frac{A_0}{\Delta x}$ values, from (2.51) and (2.52) one can find the following expression for the hydraulic conductance value:

$$K_f = \frac{A_0}{\Delta x}\frac{u^2}{8\eta}e^{6\ln^2(s)} \qquad (2.53)$$

Then, the characteristics calculated within the distributed pore model were compared with the experimental data. Discussing the above-mentioned distributed pore model, the authors summarize the obtained theoretical results as following. Notwithstanding that the model represents a simplified GBF wall within two populations (of large and small size) of pores, the theory provides a quantitative analysis of glomerular sieving and adequately describe the molecular transport across the glomerular filtration barrier.

### 3. Transport of water across natural membranes

### 3.1 Fluid pathways in the organism.

Transport of water across natural biological membranes or filters is attributed to the different mechanisms: either via special water channels called aquaporins or related to the direct transportation of water through the pores in the membrane wall.

Kidneys are natural filters in the organism. There, at the first step of the urine formation in the process of blood ultrafiltration, the crucial role belongs to the glomerulus. The glomerular capillary walls form the filtration barrier for the blood plasma sieving proteins and other macromolecular components. Damage of the glomerular filtration barrier (GFB) may cause chronical kidney disease (CKD) or lead to the kidney failure. The impared GFB function must be substituted with renal replacement therapy (RRT) or dialysis. Understanding of mechanisms of filtration involved in GFB sieving may essentially help glomerular repair therapies and dialysis improvement.

Peritoneum forms a natural biological membrane where the microvessels are distributed in the peritoneal tissues. In RRT, these peritoneal microvessels are served as dialyzing capillaries. In the peritoneal dialysis (PD), the capillaries permit slow transport of fluid and solutes. The continuous diffusive removal of small solutes and a convectional removal of large solutes from the organism across peritoneal membrane pertains a dialysis process.

The most important features of glomerular and peritoneal systems are summarized in the next two sections.

**3.2 The glomerular filtration barrier (GFB) in kidneys.**

**3.2.1. GFB structure in a nutshell**

Traditionally, the glomerular filtration barrier (GBF) was considered as a three-layer structure composed of vascular endothelial cells, glomerular basement membrane and outer epithelial cells, a podocytes foot layer (Fig.1). New experimental data revealed two more layers existence (*Arkill 2014*) – the endothelial glycocalix and the sub-podocyte space (*Salmon 2009, Arkil 2014*) (Fig.2). The role of podocytes in the GFB maintainance and the importance of glycocalix considered in details in (*Haraldsson and Jeansson 2008*). The highly debated question on the porous structure of the glomerular endothelium can be formulated as following: the glomerulum is not just a 'leaky barrier´´ (*Haraldsson and Jeansson 2008*).

In its turn, the glomerular basement membrane (GBM) comprises three fibrous layers, the inner layer (Iri), central lamina densa (Id) and outer lamina externa (Ire). The three-layer GBM provides a mechanical support for endothelial cells and serves as a molecular sieve. The latter statement has been questionable in the GFB research (*Farquhar 2006*). Barrier function of GF has been associated with slit diaphragms of podocytes. However, there are experimental evidences that the special structural features and composition of GBM also contribute to the restriction of proteins (albumin) passage (*Jarad 2006*).

The podocytes are considered as an additional filtration system. The charge selectivity properties of podocytes filter attributed to the extracellular glycocalix holding negatively charged sialic acids have been discussed in a number of recent publications (*Abrahamson and Wang 2003; Leung 2014*). It was found (*Abrahamson and Wang 2003*) that the area between foot processes of the epithelial podocytes is closed by filters or slit diaphragms. The electron tomography data display a molecular architecture of slit diaphragms as a network of protein strands containing nephrin (*Holthöfer 2007*).

In the contrast, the separating space between endothelial cells was shown contains no diaphragm. Glomerular endothelium has a fenestrated structure which also contributes to the transport of water and solutes. Simultaneously, the glomerular endothelium fenestrae are considered to be diaphragm-less, these pores seems do not efficiently restrict leakage of plasma proteins (*Abrahamson and Wang 2003*). At this point, one should bring attention to the following opinion discussed in (*Satchell and Braet 2009*). The authors (*Satchell and Braet 2009*) note, that the glomerular endothelial cell (GEnC) fenestria are similar to the filtration slits of podocytes but not studied with the same scrutiny, however may be important since the glomerular filtration rate is dependent on the surface area covered with these openings. Other topic debated in the paper (*Satchell and Braet 2009*) concerns the 'glycocalix –in-the-fenestra´´ physiological role.

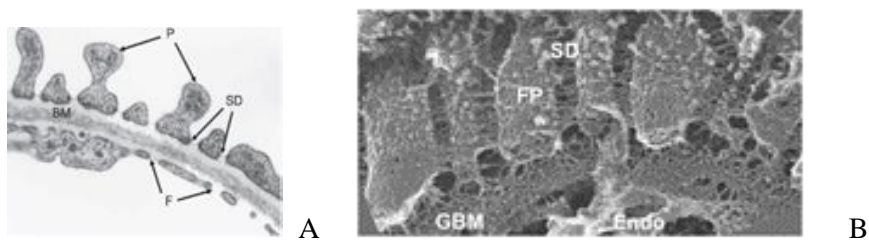

Fig.3.1A, B. Electron microscopy of the glomerular filtration barrier: P – podocytes, SD – slit diaphragms, BM – glomerular basement membrane, F – fenestrae in the endothelium (adopted from (*Satchell and Braet 2009*))

The GFB is a highly dynamic structure (*Hackl 2013*). Despite the large number of biological data on GFB, the dynamics of the filter as well as the involvement of its components and their correlated movement remain to be uncovered. The motility of the podocytes into the Bowman's capsule has been recently tracked in vivo (*Hackl 2013, Peti-Peterdi 2015*) in mouse models with multiphoton microscopy (MPM). This new imaging technique revealed a simultaneous migration of fluorescently labeled GFB cells and the spontaneous formation of cell clusters during imaging of normally functioning kidney (*Hackl 2013, Peti-Peterdi 2015*). These advanced research facilities may bring important answers on questions about what kind of mechanical motions are involved in the filtration function.

Among the most successful methods for GFB structural studies one should also mention stereology and image analysis applied for the quantification of MRI results in kidney (*Chagnac 1999, Heilmann 2011*) including glomerular number and size distribution (*Heilmann 2011*).

### 3.2.2 Modeling the glomerular sieving

The great efforts were applied to clarify sieving mechanism of sieving in glomerular filtration (see, for review, (*Layton; Edwards*; *Jarad and Miner 2009, Deen ; Comper 2008, Haraldsson 2008, Russo 2007, Chen 2008, Slattery 2008, Harvey 2007, 2008; Jeansson 2006, Greive 2001, Asgeirsson 2007, Rippe Asgeirsson 2006; Axelsson 2012, Venturoli and Rippe 2005, Öberg and Rippe 2013,2014, Rippe and Öberg 2015*), also in the review papers (*Layton; Edwards*;*Venturoli and Rippe 2005, Jarad and Miner 2009*).

Proteins and polysaccharides have been traditionally used as molecular probes for GFB selectivity studies (*Asgeirsson 2007*). It is well-established that the glomerular barrier freely filters small solutes however retaining large and negatively charged plasma proteins (*Rippe Asgeirsson 2006*). The diffusion and convection of polysaccharides such as Ficoll and dextran across the GFB is a traditional method for testing glomerular permselectivity (*Venturoli and Rippe 2005*). It was shown (*Venturoli and Rippe 2005*) that the effects of molecular size, shape, charge and deformability can also be studied with the help of these molecular probes (*Venturoli and Rippe 2005*). Glomerular sieving of a number of neutral polysaccharides has been investigated in (*Asgeirsson 2007, Rippe Asgeirsson 2006*). In comparison with proteins, polysaccharides demonstrate beneficial characteristics, for example, wider size spectrum of probes that can be used in a single experiment (*Asgeirsson 2007*). Then, the glomerular sieving coefficients obtained for these polymers can be compared with the characteristics of proteins, to resolve the effects of geometrical size on glomerular barrier permeability (permselectivity). Several findings are attracted experimental and theoretical attention, in particular, how to include the significant conformational flexibility of the molecular probes into consideration.

A distributed two-pore model suggested in (*Öberg and Rippe 2014*) has been successfully applied to the description of water and solute transport in biological membranes of different organs and recently, to artificial dialyzer membranes (*Axelsson, 2012*). In particular, the model has been applied to the analysis of the experimental data on glomerular sieving of Ficoll. The authors noted that typically, in the models applied for the filtration of solutes over a porous barrier is assumed that the molecules behave like rigid spheres. However, the experiments show that flexible polysaccharide molecules such as dextran or Ficoll, used as sieving probes, are hyper-permeable across the GBF. The calculations of the model provides the theoretical support to the idea that flexible macromolecules sieving mechanism is different in comparison with rigid spheres approach. In the model (*Öberg and Rippe 2014*), the glomerular capillary wall is represented as a barrier with two populations of small- and large-pore populations. The calculations of the model provide the theoretical support to the idea that sieving mechanism of flexible macromolecules is different in comparison with the rigid spheres approach. The variance in the distribution of pore sizes has been attributed to the molecular 'flexibility´´ of Ficoll, assuming that the true variance of the pore system is lower than the obtained using flexible probes.

The molecular probes studies allow researchers also the detailed quantitative analysis of the electrostatic effects (*Öberg and Rippe 2013*). In this approach, the glomerular filtration barrier is modeled as a charged fiber matrix that separate charged from neutral Ficoll polymers (*Öberg and Rippe 2013*). To explain the measured difference in glomerular transport between neutral and charged (anionic) form of Ficoll, the calculations of surface charge density and simulations performed for the solutes with charge density similar to that of albumin (-22 mC/m²), were carried out in (*Öberg and Rippe 2013*). The comparison of the theory with data analysis demonstrates that the electrical charge makes a moderate contribution comparing with size and conformation which seems to be more important for the filtration of macromolecules process.

**Debates on albumin sieving**. Because of the complexity of GFB structure and possible controversial interpretation of the experimental data, there is still no unified view or last judgment of how the biological components of the kidney filter contribute to the protein sieving (*Jarad and Miner 2009, Comper 2008*).

The sieving of albumin is one of the most important parameters in relation to the proteinurea characterization (*Comper 2008*). The measurable parameter is the glomerular sieving coefficient (GSC) defined as the filtrate-to-plasma concentration of the protein (*Jarad and Miner 2009, Deen 2004, Comper 2008*). The amount of filtered albumin (GSC) and the place where the protein filtration occurs has been discussed in terms of size and charge selectivity of GFB. The question - does the albumin is filtered across the glomerular barrier? Or there is some other mechanism (for example, albumin retrival hypothesis (*Russo 2007*)) that includes two parameters – size and charge, into consideration since albumin is large negatively charged protein. Discussion around albumin filtration - the glomerular vs. tubular origin, has been reviewed in a number of publications (*Jarad and Miner 2009, Deen 2004, Comper 2008, Haraldsson 2008, Russo 2007, Rippe and Öberg 2015*). In particular, in (*Comper 2008*), Haraldsson and Deen (con) strongly criticized the assumption of tubular reabsorption of proteins and mentioned that the massive experimental data provide the evidence of that the glomerular barrier normally is both size- and charge- selective. However, the defects in damaged GBF may cause the albumin leakage leading to the increased concentration of the protein. In the same paper (*Comper 2008*), another author in debates, Comper (pro), provides a support to the alternative view. Haraldsson and Deen (*Comper 2008*) also mentioned that the experiments with the radiolabeled albumin studied in (*Greive 2001*) as a proof of the albumin retrieval have been misinterpreted. Recent theoretical analysis (*Rippe and Öberg 2015*) reveals that the electrophoresis does not significantly contribute to the albumin filtration across GFB.

**3.3 Comparing glomerular and peritoneal transport**

By comparing transport across glomerular and peritoneal barriers, the relevant question arises what is the difference between these two systems?

In general, the glomerular filtration barrier displays more complex morphology than the peritoneal membrane (*Rippe Davies 2011*). Also, the highly dynamic GFB structure has been recently confirmed in (*Hackl 2013, Peti-Peterdi 2015*) studies. The uniformity of GFB is another remarkable structural feature. In this aspect, glomerulus more similar to the artificial membranes than peritoneum (*Rippe and Davies 2011*). The authors (*Rippe and Davies 2011*) noticed that due to the GFB uniformity, filtration across the glomerular capillary wall (GCW) better conforms to the pore theoretical analysis. The GFB is far less leaky than peritoneum barrier. With respect to the protein sieving, GFB is more size – selective and discriminates polymers according to their shape and flexibility (*Rippe and Davies 2011*). The GFB demonstrates also charge selectivity however this is considered now as much less influential factor.

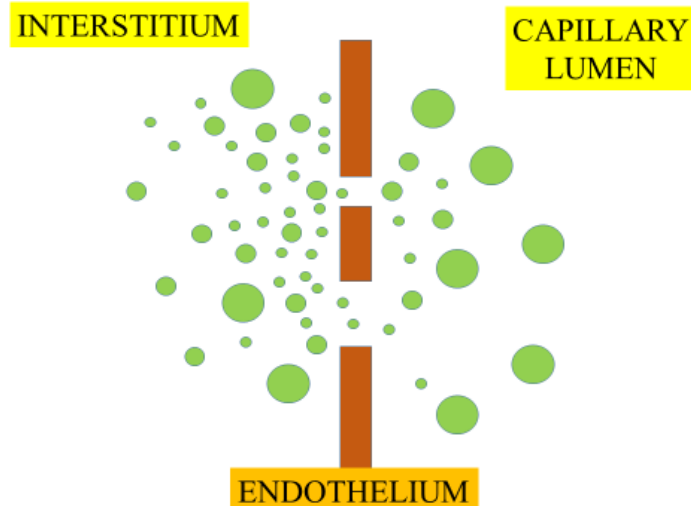

Fig. 3.2 The sketch of the peritoneal wall (after (*Rippe & Davies, 2011*)).

Morphology of peritoneum reveals the tortuosity of the interendothelial clefts and network of capillaries embedded into the interstitium. Transport across the capillary endothelium can be decribed within three-pore approach. The ultrasmall pores involved in the transport of water in peritoneal membranes are associated now with aquaporins as the 'third pores'' (*Devuyst and Ni 2006, Ni 2006, Zhang 2016*).

For the quantitative description of the transport across the peritoneal capillary wall the so-called three pore model (TPM) has been developed (*Rippe 1993, Venturoli and Rippe 2005, Rippe, Venturoli 2004*) which plays an important role for our understanding of PD mechanism.

**3.4 Three pore model (TPM) of peritoneal membrane transport**

Water transport across peritoneal membrane can be quantitatively described within the approach called a three pore model suggested by Bengt Rippe (*Rippe B, 1993*) in 90s and further developed in (Rippe 2008; *Rippe and Davies 2011, Rippe and Haraldsson 1994; Rippe and Krediet; Rippe and Levin; Rippe and Öberg; Rippe and Venturoli; Rippe and Asgeirsson etal; Rippe, Stelin, and Ahlmen; Rippe C et al*). The peritoneum which is lining out the abdominal wall inner surface in the peritoneal cavity, is a composite three –layer material of mesothelial cells, interstitial tissue containing fibroblasts, macrophages and conjunctival matrix (*Devuyst and Goffin 2008*) and a network of capillaries and blood vessels.

It is generally accepted (*Devuyst and Goffin 2008*) that during PD the endothelium lining the capillaries plays the most important role. At PD, this layer forms a barrier for water and solutes transferred from blood to dialysate introduced into the peritoneal cavity. This barrier can be described within the three pore model (TPM) as a following.

The small pores (with radius $r \sim$ 40-50 Å) between the epithelial cells fulfill about 90 % of the hydraulic conductance ($L_P S$) of the peritoneal membrane (Fig.3). The ultrasmall pores (with radius $r \sim$ 2.5 Å) in the endothelial cells contribute only 2% to the $L_P S$ value (*Devuyst and Rippe 2014*). Large pores ($r \sim$ 250 Å) viewed as interendothelial spacings which occupy less than 0.5 % of the total pore area and amount to 5 – 8 % of the hydraulic conductance of membrane (*Devuyst and Rippe 2014*).

Normally, at the absence of dialysis and without the addition of osmotic agents, approximately 60% of the transcapillary flow occurs across small pores and about 40% through large pores (*Devuyst & Rippe 2014*). It was found, that the addition of glucose as osmotic agent (SE-radius $r_{SE} \sim$ 3.7 Å), about 45% of the water flows through water-only pores and 55% across the small pores. The process of fluid reabsorption across the small pores was observed after 4 hours when the glucose gradient vanished.

Within the three pore model (TPM) approach, this partitioning of fluid flows can be attributed to the Starling forces balance as follows (*Devuyst and Rippe 2014*). In small pores, the hydrostatic pressure gradients are approximately balance each other. For large pores, the hydrostatic contribution dominates over colloid pressure since the oncotic gradients are almost negligible there.

Various osmotic agents affect the fluid removal at PD in a different way. For example, such small osmotic agents as glycerol ($r_{SE}\sim$ 3 Å) make insignificant impact to the transport across small pores while acting on water-only pores instead. In the contrast, the addition of glucose induces water flows equally partitioned between small and ultrasmall pores (*Devuyst and Rippe 2014*). During PD, the osmotic effects of glucose cause the dialysate dilution that can further lead to the reduction in sodium concentration in there (the so-called 'sodium sieving' phenomenon, which is a decreasing of the sodium in dialysate during the first two hours of the dwell) (*Devuyst and Rippe 2014*). The shift in the redistribution of Starling forces then lead to the fluid reabsorption through the small pores from the peritoneal cavity to plasma (*Rippe 2004*).

Recently *(Devuyst 2010, Zhang 2016)*, the AQP1 proteins were identified as ultrasmall pores in TPM. The water channel AQP1 studies *(Zhang 2016)* reported 'the first experimental evidence for the functional relevance of endothelial AQP1 to the fluid transport in peritoneal dialysis and thereby further validate essential predictions of the three-pore model of peritoneal transport'' (quotation from (*Zhang 2016*)). Comparison of the theoretical prediction of the TPM theory (*Rippe 1993, Venturoli and Rippe 2005, Rippe, Venturoli 2004*) with the experimental data on fluid transport across the peritoneal membrane (*Rippe 1993, Asghar and Davies 2008, Rippe and Venturoli 2008, Rippe, Venturoli 2004*) shows that this model adequately describes the pathways of peritoneal fluid transport (*Asghar and Davies 2008*). The important publication of Zhang and co-authors (*Zhang 2016*) confirms the crucial role of endothelian AQP1 in UF during peritoneal dialysis and provide a direct experimental evidence for the functional relevance of the TPM predictions. Recently (Devuyst 2010, Zhang 2016), the AQP1 proteins were identified as ultrasmall pores in TPM. The water channel AQP1 studies (Zhang 2016) reported 'the first experimental evidence for the functional relevance of endothelial AQP1 to the fluid transport in peritoneal dialysis and thereby further validate essential predictions of the three-pore model of peritoneal transport'' (quotation from (Zhang 2016)).

New updates in PD practice (*Asghar and Davies 2008, Devuyst and Yool 2010, Freida 2004*) demand further developments in the modeling. The delicate interplay between diffusive and convective processes and osmosis transport though the highly vasculated peritoneal barrier determines peritoneal dialysis conditions (*Devuyst and Yool 2010, Devuyst and Rippe 2014*). For example, more complex analysis within the three-pore model (TPM) elaborated in (*Rippe and Levin 2000*), has been employed for the prediction of the ultrafiltration profiles in peritoneal dialysis for various osmotic agents. In the next step, calculated ultrafiltration profiles (*Rippe and Levin 2000*) have been used to estimate the UF for an icodextrin (a glucose polymer, the osmotic agent used to improve fluid removal)-based PD fluid.

**Extended TPM and applications to the automated peritoneal dialysis (APD)**

In recent publication (*Öberg, Rippe 2017*) an extended 3-pore model (TPM) has been applied to the problem of optimizing patients' treatment with automated peritoneal dialysis (APD).

The APD is the process of peritoneal dialysis with the aid of a mechanical cycler with different (variable) rates of dialysate flow. The most difficult task in the APD modeling is the time-dependence of dialysis parameters at paritoneal cavity draining and filling phases in the PD cycle. The extended TPM consideration includes an additional compartment for fill-and-drain phases and combines osmotic water transport, small and middle molecules clearance and adsorption of glucose (ibid.).

Starting (at $t = 0$) from the filling phase, the net volume flow though the peritoneal membrane during APD is given by a sum of six volume terms (*Öberg, Rippe 2017*):

$$\frac{dV_{PD}}{dt} = J_{v,A} + J_{v,S} + J_{v,L} - L + J_{fill} - J_{drain} \quad (5.1)$$

Three first terms in the right hand side of equation (5.1) are the water net flow across aquaporins (index 'A''), the 'small pores'' (index 'S') considered as highly selective pathways and 'large pores'' (index 'L''), as weakly selective pathways, respectively. To include the fill (inlet) and drain (outlet) phases, two volume flows have been added. Finally, the flow $L$ represents the net lymphatic clearance from the peritoneum to the circulation (see, for the estimation of the lymphatic clearance, in (*Rippe, Stelin, Ahlmen 1986*).

In the TPM, the solute flows for each pathway have been calculated by using the Patlak equation (ibid.):

$$J_{solute,S}^i = J_{v,S}(1 - \sigma_S^i) \frac{C_D^i(0) - C_D^i \cdot \exp(-Pe_S^i)}{1 - \exp(-Pe_S^i)} \quad (5.2)$$

$$J_{solute,L}^i = J_{v,L}(1 - \sigma_L^i) \frac{C_D^i(0) - C_D^i \cdot \exp(-Pe_L^i)}{1 - \exp(-Pe_L^i)} \quad (5.3)$$

where $C_D^i(0) = C_D^i(t = 0)$, and the notations $Pe_{S,L}^i$ are used for the Péclet numbers for the small and large pores, respectively.

The kinetic equations describing the time changes in solute concentration, $\frac{dC_D^i}{dt}$ (for each $i_{th}$ solute), are given by the following formula (ibid.):

$$\frac{dC_D^i}{dt} = \frac{1}{V_{PD}}\{(J_{solute,S}^i + J_{solute,L}^i) - C_D^i \cdot (J_{v,C} + J_{v,S} + J_{v,L} + J_{fill}) + C_B^i \cdot J_{fill}\} \quad (5.4)$$

and the kinetic equations for the solutes concentration change, $\frac{dC_B^i}{dt}$, in the drain reservoir (index 'B'') are (ibid):

$$\frac{dC_B^i}{dt} + C_B^i \cdot \frac{dV_B}{dt} = \frac{1}{V_B}\{C_D^i \cdot J_{drain} - C_B^i \cdot J_{fill}\} \quad (5.5)$$

The reservoir volume $V_B$ kinetics is given by the difference in draining and filling flows:

$$\frac{dV_B}{dt} = J_{drain} - J_{fill} \quad (5.6)$$

In the formalism of thermodynamics of membrane transport (see, section 2.2 in the present review), volume flows in (5.1-5.4) can be described by using Starling formulation (ibid.):

$$J_{v,C} = \alpha_A L_P S(\Delta P - RT \sum_{i=1}^N \phi_i (C_D^i(0) - C_D^i)) \quad (5.7)$$

$$J_{v,S} = \alpha_S L_P S(\Delta P - RT \sum_{i=1}^N \phi_i \sigma_S^i(C_D^i(0) - C_D^i)) \quad (5.8)$$

$$J_{v,C} = \alpha_L L_P S(\Delta P - RT \sum_{i=1}^N \phi_i \sigma_L^i(C_D^i(0) - C_D^i)) \quad (5.9)$$

where $\alpha_{A,S,L}$ represent fractional hydraulic conductances for the different water pathways, and $\phi_i$ gives the osmotic coefficients for each of solute $i$. The reflection coefficients $\sigma_{S,L}$ are considered to be the same for the solute transport and osmotic transport *(Deen,1987)*.

By solving equations (5.1-5.6) together with (5.7-5.9) and by using appropriate initial conditions (ibid.), the authors obtained $V_{PD}(t), V_B(t), C_D^i(t), C_B^i(t)$ functions, respectively, and then, applied the results for the osmotic water transport parameters (such as ultrafiltration, UF) and molecules (urea) clearance calculations for the different dialysate flow rates (DFR). Numerical calculations based on the extended TMP model were performed in (ibid.) for $N = 25$ number of APD cycles for various DFR and in a range of glucose concentrations. The simulations of APD cycles for three different peritoneal transport regimes, slow, fast and average, allowed researchers to find the delicate balance between the glucose adsorption and the DRF and to evaluate the optimal DRF both for UF and small/middle molecules clearance.

## 3.5 Aquaporins

### 3.5.1 'Water-only'' nanopores

Aquaporins belong to the family of proteins which facilitate water transport through membranes. The involvement of these proteins in simultaneous transport of other molecules (such as $CO_2$ in erythrocytes) is controversial (*Endeward 2006, Yang 2000, Ripoche 2006, Missner 2008*).

Before the discovery of aquaporins it was assumed long time that water can freely penetrate through biological membranes. Then, family of water channels' existence has been proved: it was shown that some membranes of cells are more after permeable than other types of cells as well as pronounced osmotic reaction in comparison with diffusional water permeability. This enhanced osmosis water transportation across membranes suggested the existence of special pathways involved in water transfer. Structure of aquaporins and molecular dynamics studies revealed that water penetrates as a single-file transport manner through a nanopore in protein monomer (*Hub 2009; Verkman 2011*). The selectivity of water molecules penetration is controlled by steric and electrostatic forces.

Many biological functions of cells are regulated by the aquaporins (AQPs). The involvement of AQPs in fluid transport of cells facilitates passive water transport when large osmotic gradients are applied. It was shown that AQPs also play a key role in active fluid absorption and secretion near isosmolar processes (*Verkman 2011*). In the collecting duct epithelial cells, water transport is related to vasopresin regulation (*Verkman 2008; Noda 2010*). In kidneys' collecting duct aquaporins involved in the urine concentration mechanism facilitating osmotic water transport. In various cells AQPs facilitated water transfer across membranes plays a key role in a range of physiological functions of the cells. The selected examples are water movement involved in brain cells' swelling, cell migration, neural signaling, cell proliferation, skin hydration (aquaglyceroparins) as well as fat metabolism. The pathology in AQPs functioning may lead to the so-called human aquaporin diseases (*Verkman 2011*). The regulation of water flow in and out of cells can be influenced by extracellular osmotic disbalance and osmotic gradients across membrane. However, the influence of AQPs to the cell volume changes is under discussion now (*Verkman 2011*).

It is well established that body water balance is regulated by vasopresin. In the kidney it was found that at least 7 AQPs are expressed: AQP1, essential for urine concentration; AQP2, the predominant vasopresin-regulated water channel and others. It was discovered (*Nielsen 2002*) that vasopresin mechanism of regulation of acute water permeability in the collecting duct involves trafficking of AQP2 from intra-cellular vesicles to the apical side (toward the lumen) membrane of the epithelial cells.

### 3.5.2 Ultrafiltration related processes and aquaporins (AQP1s) regulation

Failure in the ultrafiltration (UF) is unfortunately frequent anomaly among long-term PD patients. The identification of AQP1 water nanochannel as a most relevant to UF component, brought a new understanding of transport processes across peritoneal membrane (*Nielsen 2002, Ni 2006*). Studies of AQP1s demonstrated that these ultrasmall pores in endothelial cells are essentially involved in regulation of water and ions in blood vessels of different organs (*Devuyst Yool 2010, Devuyst and Rippe 2014, Nguyen 2015*). Early investigations of swelling processes on isolated cells (*Shanahan 1999*) and on reconstituted liposomes (*Zeidel 1992*) confirmed that AQP1 proteins facilitate osmotically driven water flows. Also, the AQP1 knockout mice initial studies (*Yang 1999*) shown that the osmotically caused water transport in peritoneum of $AQP1^{-/-}$ animals was essentially diminished in comparison with wild ones (*Nielsen 2002*).

Ni and co-authors (*Ni 2006*) in mice lacking AQP1 research reported on 50 % UF decrease which was accompanied by the restriction in sodium sieving in peritoneal capillary endothelium. In the opposite, it was shown that the induction of aquaporins in the peritoneal capillaries (by corticosteroids) has stimulated water transport and UF without changes in the osmotic gradients and transport of small solutes (*Ni 2006*). These results suggest that the AQP1s are responsible for the sodium sieving and mediate half of UF during the hypertonic dwell. The regulation of AQP1 expression with the aim to increase the ultrafiltration and role of steroids has been a topic of large number of experimental efforts (see, for the review, (*Nielsen 2002, Flessner 2006, Rippe, Venturoli 2004; King 1996, Stoenoiu 2003*).

The osmoregulatory mechanisms in cells are complex and role of many factors in these processes is not well studied. The example is recently discovered coupling between the transfer of water across aquaporins and sodium transportation by molecular machines (a family of special proteins, secondary transporters) (*Zomot 2011*) across the cell membrane. The analysis of this challenging field is beyond the scope of the current review.

At the moment, there are several opinions on AQP1 role and involvement in the UF regulation. The alternative mechanism compared to the above mentioned has been reported in a number of current publications (see, for example, in (*Umenishi 2003, Belkacemi 2008, Bouley 2009*)).

**3.5.3 Aquaporins and free water transport.**

**Quantification of Free Water Transport (FWT) in peritoneal dialysis.**

The quantification of free water in peritoneal dialysis during the peritoneal equilibration test (PET) is a generally accepted method for studying solute transport and ultrafiltration in PD (*Krediet 2000, Cnossen 2009, Twardowski 1987, Krediet, Lindholm 2000; LaMilia 2005, Parikova 2005*). In the standartized PET suggested in (*LaMilia 2005*), free water transport (FWT) and small solute transport across the peritoneal barrier can be calculated after a 60 minutes dwell with 3.86% glucose by analyzing the sodium transport kinetics. (For the detailed description of the standard PET procedure see, for example, paper (*Cnossen 2009*)). For the FWT calculation, simultaneously, the intraperitoneal volume should be measured (*Krediet 2000, Cnossen 2009, Smit 2004, LaMilia 2005*). In the peritoneal equilibration test, the sieving of sodium which is considered associated with a hypertonic glucose solution allows researchers to value aquaporin-related transport of water (*Rippe 1991*). The UF is defined as a ratio of the transported sodium amount to the plasma sodium concentration (*Cnossen 2009*). The FWT is calculated as a difference between the total UF after one hour and the small-pore transport value (*Cnossen 2009*):

$$FWT = UF_{Total\ after\ 1\ hour} - UF_{small-pore\ transport}$$

During PD, crystalloid osmotic pressure gradients induce transport of fluid which incorporates transport across small pores and free water transport (*Krediet, Lindholm 2000; Parikova 2005*). The FTW is considered as related to the aquaporin AQP1s pathways (*Ni 2006*). The input of both pathways can be quantitatively determined by comparing the kinetics of water and sodium transport (*Smit 2004, LaMilia 2005, Krediet 2000, Cnossen 2009*).

Fluid pathways and reabsorption in the peritoneum within the TPM concept have been experimentally studied in (*Asghar and Davies 2008*) at lower concentration of glucose than suggested by LaMilia and co-authors in (*LaMilia 2005*). By using the radio-labeled albumin as an intraperitoneal volume marker and low concentration (1.36%) glucose solution, Asghar and Davies (*Asghar and Davies 2008*) determined changes in the intraperitoneal sodium concentration and, then, evaluated the transperitoneal clearance of sodium (*Asghar and Davies 2008*). This work deserved a comment of Rippe and Venturoli (*Rippe and Venturoli 2008*) on back-filtration of fluid through the small pores in view of the three pore model.

**3.5.4 Debates**

Physiology of transperitoneal exchange received close attention and numerous discussions around free water transport mechanisms and role of aquaporins (*Nielsen 1993, 1999, 2002, Kishore 1996, Me and*

*Taylor 1989, Knepper 1997, Loffing 2000, Ni 2006, Aguirre 2014, Flessner 2005, 2006, Rippe 2008, Ni 2006*). In particular, Flessner in the comment (*Flessner 2006*) to the paper by Ni and co-authors (*Ni 2006*) discussed the potential influence of the glycocalix to the peritoneal transport. The author also considered a role of endothelial glycocalix in abnormal cells and argued that the defects in damaged glycocalix could form pathways for the fluid and solutes transport. The author recollected the so-called 'pore-matrix theory'' (*Fu 1995*) and referred to the experiments (*Henry 2000*) where it was shown that the inflammation can lead to the glycocalix alteration. The latter, according to the author's opinion, may increase the permeability of the endothelium to solutes such as glucose and sodium that, in its turn, influences the UF.

### 4. Water transport across the artificial dialysis membranes.

The artificial kidney substitutes the natural renal function with the process of blood purification in the hemodialysis procedure using filter membranes of various chemical composition and microstructure (*Ronco, 2004*).

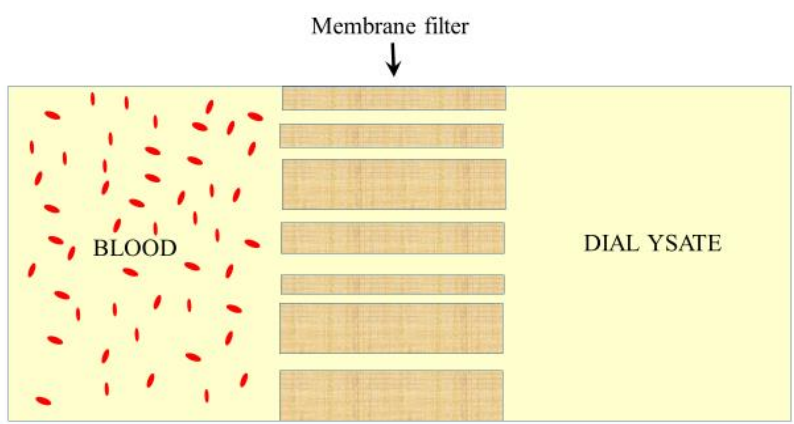

Fig.4.1 Schematics of hemodialysis filtration

The main goal of the improved filtration in hemodialysis (HD) is related to the design of artificial polymer membranes for the dialysis capillary filters with the special properties. This general perspective involves several aspects. First, the technological aspects require polymers with special physical and chemical properties to fabricate non-woven networks and/or porous membranes. The experimental studies of the filter membranes includes the ultrastructure analysis (electron microscopy) and mass transport across membranes. Design of better HD filters requires mathematical methods of modeling and computer simulations to study and analyze pore size and geometry, including topological aspects, connectivity of pores and tortuosity factor. New advanced mathematical tools such as stereology (*Heilmann 2011*) can be used for the predictive modeling in nanoporous filter material design.

Since the HD blood purification demands multidisciplinary efforts, both engineering improvements and biomedical search for the optimal filters are needed to be developed in parallel with clinical studies of new synthetic materials.

### 4.1 Microporous membranes for hemodialysis: ultrastructure of pores and their characteristics

There are special requirements for membranes in dialysis such as adequate diffusion and convection characteristics, sieving coefficients, cut-off point similar to glomerulus (*Ronco 2004*), adequate ultrafiltration rate, suitable biocompatibility properties, non-toxicity of the material, as well as the non-degradable stability for the sterilization (which may affect physical properties of the filters) and demand to be re-used searching the low-cost solutions (*Ronco 2004*). Also, reproducibility of dialysis and constant performance during entire procedure (*Ronco 2004*).

Biocompatibility of polymer membranes is an important concern in HD applications. Biophysics and biochemistry of the membrane biocompatibility (blood compatibility) among the hardest pieces of work and basic relations still remain to be uncovered.

Hollow polymer capillaries (Fig.4.1A, Fig.4.2, Fig.4.3) for ultrafiltration despite the variations in chemical composition[1] and microscopic details of pores share the common features in their structure. Their 3-layer membranes belong to the ISA (Integrally Skinned Asymmetric) (*Marchetti 2014*) class of filter membranes. The ultrastructure of the ISA wall cross-section is schematically shown on Fig.4.1B.

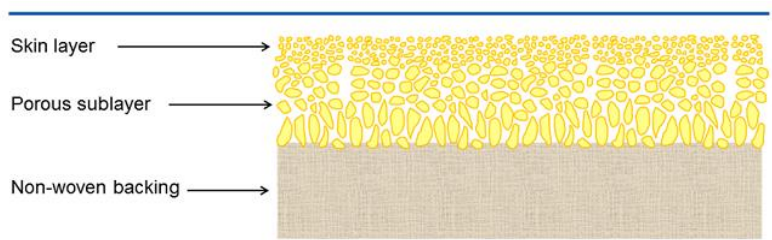

Fig.4.1. Schematics of the three layer structure of filter membrane (adapted from (*Marchetti 2014*))

In the integrally skinned asymmetric (ISA) membranes, the inner surface of a capillary is a skin layer of dense polymer with nanometer pores. The skin layer borders with the highly porous sublayer, a 'spongy'' material of large voids separated with polymer fibers (Fig.4.1). Non-woven backing layer is separated from the skin layer with a highly porous sublayer. The outer surface of backing layer is in the contact with a dialysate. The structure of this layer is different from the skin layer and porous sublayer. The latter one is a 'spongy'' material of large voids separated with polymer fibers.

**4.2 Characterization of membrane pore geometry in the capillary filters. Ultrastructure of porous membranes. Fouling of membrane filters.**

Electron microscopy of capillary wall of HD filters is a common tool for the morphological characterization of membrane porous structure. The structure of pores varies from layer to layer. Numerous research papers reported the ultrastructure of the inner 'sponge'' membrane area for different type of polymers.

Hollow polymer capillaries for hemodialysis despite variations in the composition and microscopic pore details, share the common features in their structure.

---

[1] Common polymer materials for hemodialysis membranes are polysulfone (PSf), polyethersulfone (PES), polyamide (PA) and cellulose acetate (CA) (High performance membrane dialyzers. Editor(s): Saito A., Kawanishi H. , Yamashita A.C., Mineshima M.  Series Contributions to Nephrology).

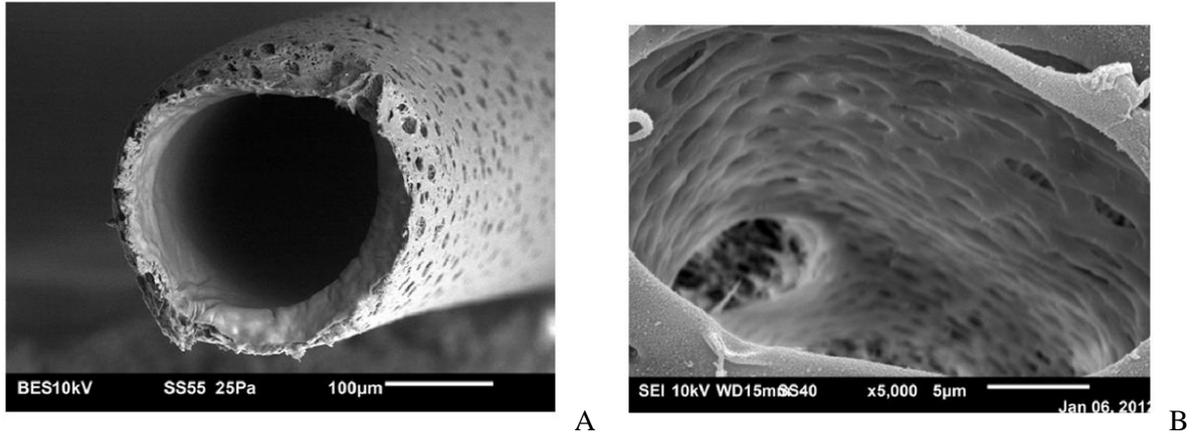

Fig.4.2. A - SEM of the cross section of a Polyflux capillary at 250x; B – SEM of a Polyflux capillary wall showing tortuous pore morphology (outside pore in contact with the dialysate) at 5000x (adopted from *(Hedayat A., J Szpunar 2012)*)

The skin layer of the capillary is a blood-contacting surface and a primary zone for molecular sieving during the UF process. The inner surface skin layer has a tight structure penetrating with nanometer-sized pores. The chemical composition of polymer and the porosity of the membrane determine the membrane permeability for water and solutes. Since during HD the blood stream, perpetually washing the capillary walls, brings cells and plasma proteins in contact with the skin layer, the complex interaction of these blood components with polymer is accompanied with a number of clinical problems. The geometry of nanopores in the skin layer depends on chemical composition of polymers and method of membrane preparation. Figs.4.2 -4.3 show EM of microstructure of two polymer membrane filters used for hemodialysis, Polyflux 201H HD membrane (*Hedayat 2012*) and polysulfone (PSf) capillary (*Repin* etal)).

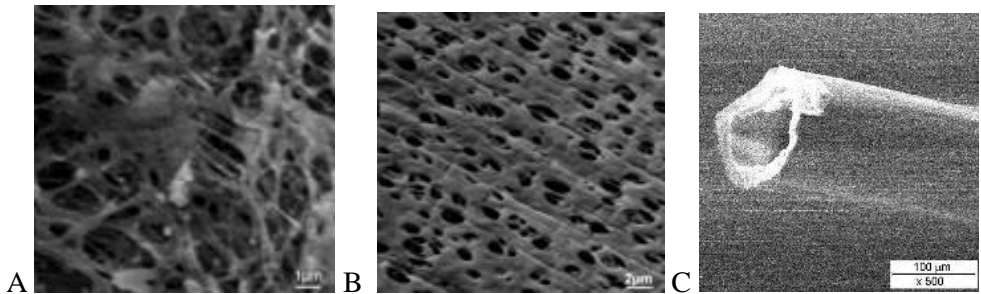

Fig.4.3. SEM of the polysulfone (PSf )capillary ultrastructure (a dissected HD capillary wall) : A – SEM of the sublayer voids (a sponge-like membrane area); B – SEM of the outer surface of PSf capillary (in contact with the dialysate ) (*Repin 2014-2015*); C - view of the PSf hollow capillary cross section (*SEM Petr Savitcki, assisted D.Gaman, MV*)

Further improvement in the material science of membrane filters for dialysis includes essential efforts in prevention of the biofouling and cake layers formation. Search for new blended polymers is a rapidly developed branch in physics and chemistry of polymers. Another direction is a surface functionalization of dialysis filters with macromolecular compounds with antithrombotic properties and polymers reducing membranes fouling.The surface functionalization of the membranes may essentially change the architecture of its skin layer (Fig.4.4). The experimental examples of surface functionalization is provided by grafted polymers coating of dialysis filters (*Rana and Matssura 2010)* and composite carbon nanotube - PES membranes with heparin-mimicking polymer brush (*Nie etal 2015*).

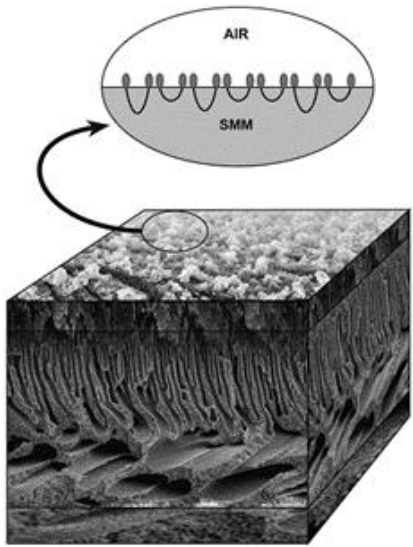

Fig.4.3.Ultrastructure of the skin layer of filter membrane after its surface functionalization with chemical groups on the top (adapted from (*Rana and Matssura 2010*)).

Due to the competitive adsorption of high molecular weight proteins (such as albumin, fibrin, fibronectin or globulins, and to the second step, adsorption of low and medium molecular weight proteins such as beta2-microglobulin, cytokines), the formation of cake ad-layer which forms a barrier for the transporting fluid occur. The structure of the formed cake layer –random or ordered- determines the pores geometry (straight channels, randomly distributed voids or tortuous pathways) which may influence the transport of fluid across the cake layer in similar way to the filter membrane by itself. However, within the cake layer, not only its geometrical and topological characteristics play role but also the dynamics of enzymatic reactions involved in the construction-deconstruction of the cake layer.

A special attention has been paid to heparin which is widely used as an antithrombotic substance in dialysis circuits. Permeability of dialysis membranes for water and solutes has been experimentally studied for the pristine as well as for the surface functionalized filters. In particular, for the heparin coating (of cuprophane membranes), the enhanced water permeability was found (*Hinrichs 1997*) in comparison with untreated membranes. However, heparinization of dialysis filters remains a debated area (*Shen and Winkelmayer 2012*).

Theoretical consideration of the fouling and ultrafiltration process in the cake layers makes an additional, complementary part in the quantification of complex dialysis processes. Next section represents the short review of selected theoretical models describing hindered diffusion through the synthetic polymer membranes used for dialysis.

5. **Mathematical modeling of molecular transport across synthetic dialysis membranes**

**5.1 Size-selectivity of a synthetic dialysis membranes**

In addition to the two-pore model, discussed in the section 2, a heteroporous model (*Rippe 2006*) with a log-normal distributed population of pores in parallel with a non-selective shunt was used to validate the experimental filtration data (*Axelsson, Öberg 2012*). Within the so-called $\theta$-model, the theoretical $\theta$ data were calculated from the non-linear convection/diffusion equation derived. Under assumption of a non-selective shunt, the $\theta$-model yields:

$$\theta_{model} = f_D \frac{1-\sigma}{1-\sigma e^{-Pe}} + f_L \qquad (5.1)$$

where the Péclet number ($Pe$) is defined as following:

$$Pe(r_S, GRF) = f_D \frac{GFR(1-\sigma)}{k \cdot A} \qquad (5.2)$$

Here $k$ is the membrane permeability, $A$ is the total pore surface area, $\sigma$ is the Staverman reflection coefficient, $f_L$ is the fractional fluid flow through shunts ($f_L = 1 - f_D$).

One should notice that in the $n$-pore model, pore radii are assumed to be *discretely* distributed respective to their weights, $\alpha_i = L_{P,i}/L_P$. The total reflection coefficient for $n$-pore membrane is given by the expression:

$$\sigma_n(r_S) = \sum_{i=1}^{n} \frac{L_{P,i}}{L_P} \sigma_h(r_S, r_i) \qquad (5.3)$$

In the expression (5.3), the hydrodynamic reflection coefficient:

$$\sigma_h(r_S, r_i) = 1 - \frac{(1-\lambda)^2 (2-(1-\lambda)^2)(1-(\lambda/3))}{1-\left(\frac{\lambda}{3}\right)+\left(\frac{2\lambda^2}{3}\right)} \qquad (5.4)$$

where $\lambda = r_S/r_i$ is the ratio of solute radius to the pore radius, respectively.

Simultaneously, in the distributed model, it is assumed that the pore radii are *continuously* distributed according to the ordinary log-normal distribution:

$$f(r) = \frac{1}{\sqrt{2\pi} r \ln s} \exp\left(-\left(\frac{\ln r - \ln u}{\ln s}\right)^2 / 2\right) \qquad (5.5)$$

In formula (5.5.) $u$ denotes the mean pore radius and $s$ is the distribution width.

The hydraulic conductivity of the porous media with the distribution of pores of a fixed radius $R$ can be calculated by using Poiseuille's law

$$L_{P,R} = \frac{\pi R^4}{8\eta H} f(R) \qquad (5.6)$$

where $\eta$ is the viscosity of water, $H$ is the membrane thickness.

For the distributed model, the total reflection coefficient is given by the expression:

$$\sigma = \int_0^\infty \frac{L_{P,i}}{L_P} \sigma_h(r_S, r_i) dr = \frac{\int_0^\infty r^4 f(r) \sigma_h(r) dr}{\int_0^\infty r^4 f(r) dr} \qquad (5.7)$$

Then, to evaluate the size-selectivity of dialysis membranes, the theoretical $\theta$ values obtained in the model were compared with the experimental data for Ficoll (FITC) filtration.

A new development in this field includes the effects of membrane fouling. Current work of Polyakov and Zydney (*Polyakov, Zydney 2013*) combines a complete blocking model (see, for the classical complete blocking theory, papers (*Hermia 1982, 1985*)) and the approach based on thermodynamics of membrane transport. In the framework of thermodynamic description (section 2) extended to the case of diffusion across the fouling layer, the volume flux can be written as follows:

$$J = \frac{\Delta P - \sigma \Delta \Pi}{r_M + r_F} \qquad (5.8)$$

where $r_M$ and $r_F$ are the hydraulic resistance (i.e., the inverse hydraulic conductance) of the membrane and of the fouling (cake) layer, respectively (*Polyakov Zydney 2013*). Based on the formulation of (*Santos etal 2006,2008*), further model calculations permit analysis of two pores sizes and permeate fluxes as well as calculation of solute rejection coefficients (see, for details of calculations, in ( *Zydney 2011; Polyakov & Zydney 2013*)).

Recent publications on modeling of UF membranes performance based on similar theoretical framework have been reviewed in (*Polyakov and Zydney 2013*). In particular, the effect of pore size distribution on permeability of membranes has been studied in (*Zydney 2011; Mehta and Zydney 2005; Kanani Fissell 2010; Mehta Zydney 2006).* The influence of pore size distribution and pore connectivity distribution has been studied for the diffusive transport in model porous networks by Armatas (*Armatas 2006)* and co-authors (*Armatas, Salmas, 2003*). The results of calculations proved that the pore size distribution and related percolation phenomena essentially affected the tortuosity and diffusivity of the porous network (*Armatas 2006*).

Modern polymer technologies allow fabrication of synthetic membranes with predetermined geometry of pores (see (*Kanani 2010, Mehta and Zydney 2005*)). Model silicon membranes are convenient objects for the analysis of the effects of pore size and pore geometry. In their recent study Kanani, Fissel and co-authors (*Kanani 2010*) compared the permeability of silicon membranes with pores of slit-shaped and cylindrical geometry. The result of calculations has shown that the membranes with slit pores demonstrated a higher performance (i.e., for a given permeability showing higher selectivity) than ones with cylindrical pores (*Kanani 2010*). The improved performance, however, was found became lesser when the pore size distribution increased (*Kanani 2010*). These results demonstrate the complex interrelation between the effects of pore geometry and size distribution and thus, provide the new insight in the ultrafiltration process.

### 5.3 Modeling diffusion across the tortuous membrane pathways

#### 5.3.1 The 'hydraulic tortuosity'' notion.

The complexity of the membrane ultrastructure visualized electron microscopy, possess a challenging task to include in the model not only geometrical size of pores but also the topology of the porous space. The microscopic pores topology has been suggested to characterize in terms of 'interconnectedness of their shapes,'' such as 'connectivity'' and 'tortuosity factor''. (*Dullien 1992*). Notion 'tortuosity'' it is customary to apply to the analysis of fluid flows across granular beds (*Kozeny 1927, Carman 1937*) or porous media (*Dullien 1992, Armatas 2006, Armatas Salmas 2003, Kim and Chen 2006*).

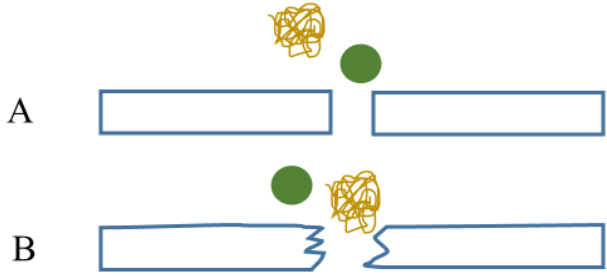

Fig.5.1.. Schematic depiction of the membrane pore : A – cylinder channel pore; B – tortuous pore

Tortuosity factor was introduced in the fundamental work of Carman (*Carman 1937*) as the square ratio of the effective average path length, $L_{eff}$, to the shortest distance, $L$, along the fluid flow direction (*Dullien, 1992*):

$$\tau = \left(\frac{L_{eff}}{L}\right)^2 \qquad (5.9)$$

In the later models, the notion of the 'hydraulic tortuosity'' of porous media filled with a liquid, has been suggested (*Dullien 1992*). The 'hydraulic tortuosity' factor was introduced as the ratio of the cross section available for conduction (flow) to the bulk cross section which was taken be equal to the bulk porosity $\varepsilon$. The total 'pore volume'', $V$, can be defined as (*Dullien, 1992*) $V = \varepsilon L^3$, respectively. In the next section the derivation of the expression (5.9) is provided by using the analogy between the electrical conduction and diffusion in porous media (see, the subsections 5.3.2-5.3.4, and the 'Formation resistivity' in particular).

In various models, volume porosity value $\varepsilon$ has been related to the fluid flow characteristics. For example, the expression for the averaged velocity in the porous media flow channels (or 'pore velocity') in Hagen-Poiseuille type equation is given by the formula (*Dullien 1992*):

$$v_{inside-the-pore} = \frac{\Delta P}{L_{eff}} \cdot \frac{D_h^2}{16 K \eta} \qquad (5.10)$$

where $\Delta P$ is the hydrostatic pressure, $D_h$ is the 'hydraulic diameter', $K$ is the 'shape factor', $\eta$ is the shear viscosity of fluid. The hydraulic diameter was shown (*Dullien, 1992*) can be related to the specific surface area of solid volume, $A_{solid}$, and porosity $\varepsilon$ as following:

$$D_h = \frac{4\varepsilon}{A_{solid}(1-\varepsilon)} \qquad (5.11)$$

Usual form of the Carman-Kozeny equation for the permeability coefficient $k \equiv k_{CK}$ is given by the relation:

$$k_{CK} = \frac{\varepsilon^3}{K\tau^2(1-\varepsilon)^2 A_{solid}^2} \qquad (5.12)$$

## 5.3.2 Analogy between electrical conduction and diffusion in porous media

One of the most useful physical parallels in analysis of mass transfer in porous materials is the analogy between diffusion in porous media and electrical conductivity (*Dullien 1992; Klinkenberg; Wyllie 1957*).

The conductivity of a porous medium (proportional to the porosity, $\varepsilon$) is often described by using a resistivity (or formation factor, $\Phi$) introduced as an inverse to $\varepsilon$ value, however, it was found that $\Phi(\varepsilon)$ has more complicate form (*Dullien, 1992*). In particular, by using the analogy with the electrical characteristics of non-conducting (solid phase) particles embedded into conductive medium (fluid inside the pores), one can use the expression for the conductivity (or resistivity) of the mixture. For example, for the uniform non-conductive spheres in a bulk fluid, the Maxwell model (*Maxwell 1881*) provides the following expression relating the resistivity (formation factor) $\Phi$ and porosity, $\varepsilon$:

$$\Phi = (3 - \varepsilon)/2\varepsilon. \qquad (5.13)$$

The electrical-fluid flow analogy and Maxwell's expression (5.13) have been used in many studies of flows through granular materials, in particular, for the analysis of transport in tortuous solid (colloidal) and soft (biofilm) cake layers (see, for the review, *Kim and Chen 2006*).

In the next section it is shown how in a simple scheme the formation factor $\Phi$ and porosity $\varepsilon$ can be related to the geometrical parameters of a porous medium or its tortuosity factor $\tau$.

The important physical clarification about the electrical conductivity and diffusion analogy should be introduced here. For the consistent view, let us consider conductivity of the anisotropic material (*Landau, Lifshitz 1960*). In general, the relation between the electrical current density $j$ and the electrical field $E$ is:

$$j_i = \gamma_{ik} E_k \qquad (5.14)$$

where $\gamma_{ik}$ is a symmetric tensor of conductivity.

The symmetry of conductivity tensor

$$\gamma_{ik} = \gamma_{ki} \qquad (5.15)$$

follows from Onsager's principle of symmetry of kinetic coefficients. (This fundamental principle has already been mentioned in the section 2 of current review, see the derivation of Kedem-Katchalsky kinetic equations).

In general, by introducing dynamic variables describing at each point the state of the system $\{\chi_n\}$, the velocities $\frac{\partial \chi_n}{\partial t}$ and corresponding generalized forces, $\{X_n\}$, one can write (Landau, Lifshitz, 1960) for the time variation of total entropy $\Theta$ of the system:

$$\frac{d\Theta}{dt} = -\int \sum_n X_n \frac{\partial \chi_n}{\partial t} dV \qquad (5.16)$$

In particular, for the time variation of the entropy of the anisotropic conductive body in electric field one can get:

$$\frac{d\Theta}{dt} = \frac{1}{T} \int (\vec{j} \cdot \vec{E}) dV \qquad . \qquad (5.17)$$

$T$ denotes temperature, $V$ is the body volume.

By comparing (5.14), (5.16) and (5.17), one can easily establish the correspondence between electrical current density components and velocities (*Landau, Lifshitz 1960*):

$$\frac{\partial \chi_n}{\partial t} \to j_n \qquad (5.18)$$

and between the generalized forces and components of the electric field

$$X_n \to -\frac{E_n}{T} \qquad . \qquad (5.19)$$

From above mentioned, one can see, that linear equations describing the electrical conduction and diffusion in liquid phase have a similar structure and obey Onsager's principle of symmetry of kinetic coefficients. This similarity has a fundamental consequence which is considered below in the subsections 5.3.3-5.3.6.

The electrical conduction – diffusion analogy has been extensively used for the theoretical analysis of transport of molecules through porous media and granular materials (*Carman; Klinkenberg; Dullien*).

### 5.3.3 Formation resistivity factor and tortuosity.

The kinetic equations describing diffusion in a liquid and electrical conduction are linear relations between the thermodynamic fluxes and conjugated forces, respectively. In an open space filled with liquid, these equations are written in the form of Fick's law and Ohm's law for the mass flow $j_m$ and current density $j_e$ :

$$j_m = -D\nabla n \; , \qquad (5.20)$$

$$j_e = -\gamma \nabla U \; , \qquad (5.21)$$

where $D$ and $\gamma$ are the diffusion coefficient and the specific electrical conductance, respectively, $\nabla U$ is the electrical potential gradient.

Diffusion of solutes (for example, trace molecules) through a fouling (cake) layer can be described in terms of hindered diffusion in a porous medium. For the hindered diffusion of solutes in a porous fouling layer of porosity $\varepsilon$, one can rewrite the equation (5.20) as follows (*Kim and Chen*):

$$j_H = -\frac{D_0}{\Phi}\nabla n = --D_H \varepsilon \nabla n \qquad (5.22)$$

where $\Phi$ is the formation factor, $D_H$ is the hindered diffusion coefficient of solutes in the fouling (cake) layer and the tortuosity factor $\tau = \frac{D_0}{D_H}$ is given by the ratio of the diffusion coefficients, respectively (*Kim and Chen*).

In a fouling (cake) layer filled with liquid, solute molecules diffuse across conducting capillaries in a solid matrix. Here, the effective diffusion coefficient and electrical conductivity become a function of two factors characterizing the medium: its porosity ($\varepsilon$) and the shape of the conductive capillary (pore) or tortuosity. According to Dullien, tortuosity of porous materials is a fundamental property which measures the deviation from the macroscopic flow at every point of fluid (*Dullien*). In general, tortuosity is a tensor value (*Dullien, 1992*). However, for isotropic material the tortuosity tensor reduces to a scalar (a tortuosity coefficient, $\tau$) (*Dullien 1990, 1992*).

The relation between tortuosity of the fouling layer, its porosity and the resistivity formation factor can be clarified by using the following example.

Porous medium geometry in its simplest form can be represented as a uniform subset of conductive capillaries (channels in a solid matrix) having the same length but varying in diameters (Fig.5.2) (*Dullien, 1990, 1992*).

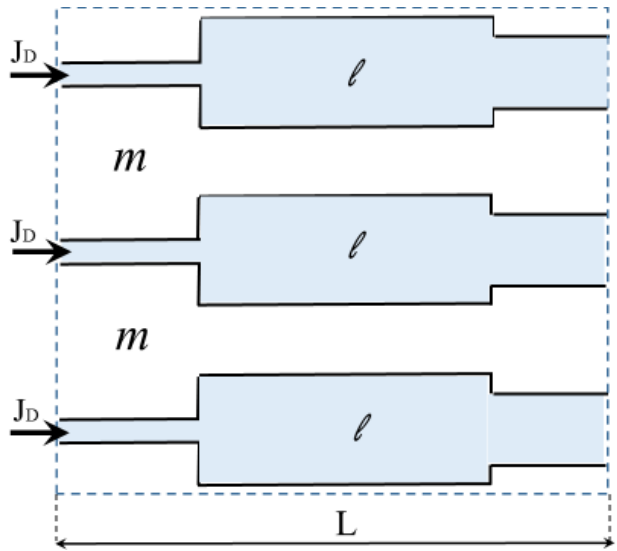

Fig.5.2. A schematics of porous material represented as a composite of solid matrix (index $'m'$) and tortuous channels filled with liquid (index $'l'$).

For a porous material depicted on (Fig.5.3), the tortuosity $\tau$ is given by the geometric ratio:

$$\tau = \left(\frac{L_l}{L}\right)^2 \qquad (5.23)$$

and the material porosity $\varepsilon$ (which is a ratio of the void space and solid matrix volume) is expressed as follows (*Dullien, 1990*):

$$\varepsilon = \frac{L_l a_l + L_m a_m}{A(L_l + L_m)} \qquad (5.24)$$

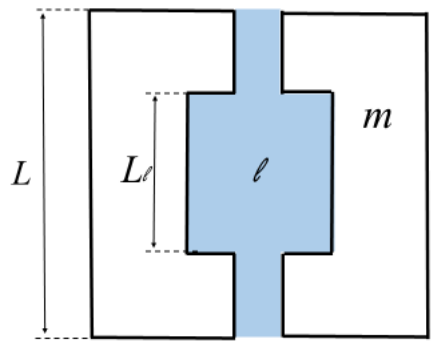

Fig.5.3. The schematics of actual pore geometry of length $L$ in the uniform capillary model

Besides porosity and tortuosity, another parameter called the resistivity formation factor, $\Phi$, is equally widely used. According to (*Wyllie, Dullien*), the formation resistivity factor can be introduced as the ratio of the electrical resistance $R_p$ of the porous material filled with an ionic solution, to the bulk resistance $R_l$ of this solution in a volume occupied by porous space, $\Phi = R_p/R_l$. This value provides a measure to evaluate the influence of porosity on the electrical resistance of material. In the model approach of (*Wyllie; Dullien*), the resistivity formation factor and porosity were related as following:

$$\Phi \equiv \frac{\Upsilon}{\varepsilon} \qquad (5.25)$$

where $\Upsilon$ is 'electrical tortuosity', which is given by the ratio of molecular diffusivity to the effective molecular diffusivity (*Dullien*). For the porous sample presented on (Fig.5.2), the resistivity formation factor $\Phi$ is given by the ratio:

$$\Phi = \frac{R_p}{R_l} = \frac{\rho_l(L_m/a_m + L_l/a_l)A}{\varrho_l(L_m + L_l)A} \qquad (5.26)$$

By using formula (5.25) and the relation $L = L_m + L_l$, one can find from (5.26) that

$$\Phi = \frac{1}{\varepsilon}\left(\frac{L_l}{L}\right)^2 = \frac{\tau}{\varepsilon} \qquad (5.27)$$

By comparing (5.23) and (5.27), one can notice the equivalence of these expressions, so for the geometry of (Fig.5.3), $\Upsilon = \tau$.

For more complex geometry of the channels, a generalized expression for the resistivity formation factor was introduced:

$$\Phi \equiv \frac{\tilde{\Upsilon}}{\varepsilon},$$

$$\tilde{\Upsilon} = \tau \cdot \tilde{S} \qquad (5.28)$$

where $\tilde{S}$ is the 'constriction factor' ($\tilde{S} \geq 1$) (*Dullien, 1992*).

### 5.3.5 The geometrical model of the hindered diffusion in fouling layer composed of microspheres

Neale and Nader (*Neale and Nader*) formulated a geometrical model which provides a deep physical insight into diffusion processes in fouling layers. The authors analyzed the transport in a porous medium, a spherical cavity in an homogeneous isotropic swarm composed of microspheres of different size.

In the geometrical model suggested in (*Neale and Nader*), the ratio of radii (Fig.5.4):

$$\frac{R_0}{R_1} = (1 - \varepsilon)^{-\frac{1}{3}} \qquad (5.29)$$

where $\varepsilon$ is porosity of the material introduced in the model as a ratio of outer shell - to the reference sphere volumes:

$$\frac{v_0}{v_1} = 1 - \varepsilon \qquad . \qquad (5.30)$$

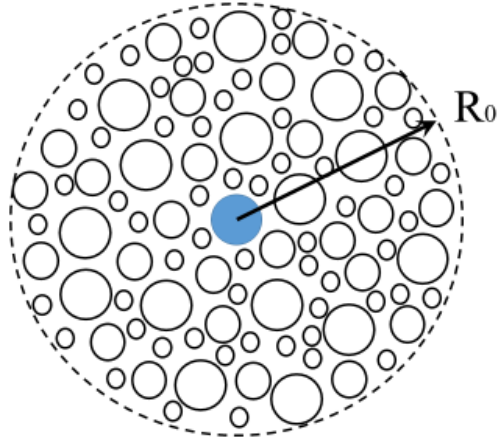

Fig.5.4. A sketch of homogeneous and isotropic swarm of microspheres of porosity ε illustrating the geometrical model (from (*Neale, Nader*). Figure shows a cross section through the center of reference microsphere (marked with blue colour) of radius $R_1$ and the outer sphere of radius $R_0$ (the concentric shell, marked with dotted line).

By solving kinetic transport equations

$j_{10} = -D_1 \nabla n_1$ (5.31)

$j_{0\infty} = -D_0 \nabla n_0$ (5.32)

with the appropriate boundary conditions within the spherical shell ($R_1 < R < R_0$) and within the exterior porous material ($R_0 < R < \infty$), the authors calculate corresponding macroscopic fluxes $j_{10}, j_{0\infty}$ and evaluate the diffusivity factor $\Lambda = \frac{D_0}{D_1}$ ( defined as a ratio of (effective) diffusivity in the porous medium to the absolute diffusivity, i.e. diffusion in the fluid without obstacles):

$\Lambda = \frac{2\varepsilon}{3-\varepsilon}$     $(0 \leq \varepsilon < 1)$     (5.33)

In fact, the diffusivity factor $\Lambda$ used by the authors (Neale, Nader) is the inverse formation factor $\Phi$, so the result (5.33) reduces to the Maxwell's formula (5.13). The authors emphasized that the boundary value electrical conduction of mixture problem (*Maxwell; Wagner; Rayleigh, Bruggemann; De la Rue and Tobias*) and the diffusion problem solved in the proposed geometrical model (the interstitial diffusion) are mathematically equivalent.

### 5.3.6 Diffusive tortuosity and random walks simulations

Further development of the fundamental theory presented above the interested reader can find in the current works on molecular diffusion in porous materials (see, for example, in (*Hizi and Bergman 2000); (Mitra, Sen and Schwartz,1993* ). In (*Hizi and Bergman, 2000*), the authors suggested a theoretical model for the diffusion of fluid in porous media with periodic microstructure. In the model, a diffusion time-dependent coefficient $D(t)$ is introduced. The calculation of $D(t)$ value shown that this coefficient is dependent also on the absolute size scale and on topology of the porous microstructure. In this work, for the calculation of the effective diffusion coefficient, the analogy of the stationary diffusion and the electrical conduction in the pore space has been used (for the insulating matrix, $\gamma_m = 0$).

Diffusion tortuosity factor of solid and soft fouling (cake) layers has been investigated in current theoretical work (*Kim and Chen, 2006*). In this publication, the random walk simulations of solute traces have been done for the different geometries of the porous cake layer for the periodic as well as random pathways in the cake layer (Fig.5.5).

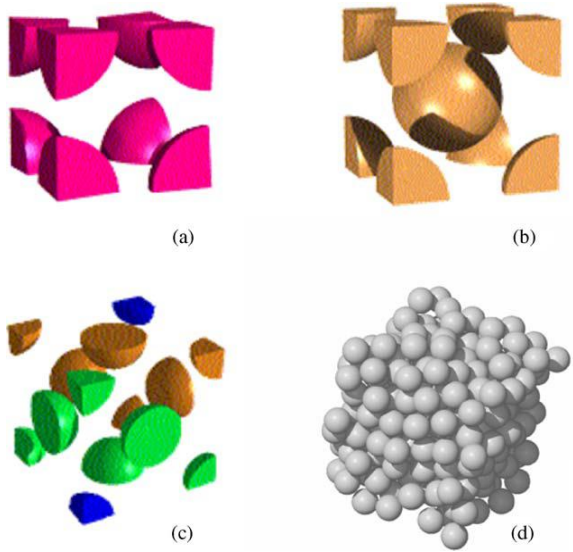

Fig.5.5. Schematics of a cake layer with periodic (a, b, c) and irregular structure (d, random colloidal structure of interacting monodispersed spheres) (adopted from (*Kim and Chen, 2006*)).

Within the hindered solute diffusion theory, the authors modeled fouling and ultrafiltration in colloidal porous cake layers (*Kim and Chen, 2006*). In particular, by using Maxwell's formula (5.13), the authors (*Kim and Chen*) related the diffusive tortuosity factor with the cake volume fraction and calculated corresponding diffusive flows.

## 6. Discussion and outlook

Water transport and filtration in living organisms are processes undoubtedly crucial for their normal functioning and metabolism. A failure of kidney, the main filtering system of the organism, changes the total water balance resulting in over- or dehydration of tissues, wrong redistribution of chemical components and general intoxication of the organism. The abnormalities in water and solute transport are also related to a number of diseases affecting circulatory, cardiovascular, and neuronal systems as well as and lung diseases. The concentration of proteins in the interstitial body fluid, lymph and blood plasma influences the oncotic (or colloid osmotic) pressure in these main bath solutions so the protein outbalance may lead to unwanted water gradients across the cell walls in vessels, tissues and organs. For example, the reduced level of albumin, a major protein in blood plasma, as well as other proteins diminish the oncotic pressure in the blood, possibly leading to edema formation (*The Kidney: Physiology and Pathophysiology, 2007*). The concentration of saccharides, most notably glucose, is essential for the maintenance of energy and healthy status of the body. The transportation of water along the fluid pathways (vessels, capillaries) bringing cells and nutrition to the body organs and across the cell wall are the physiological processes where the hydrostatic and osmotic (and colloid osmotic) pressure play a key role.

In the natural filtering systems of the organism, the separation or sieving of small solutes, proteins, and saccharides occurs in the multilayer biological membranes characterized by special composition and enormous complexity in architecture. When the natural biological filters of the organism are damaged, the artificial dialysis systems are employed to repair the lost sieving function.

Two common treatment procedures for patients with renal failure are hemodialysis (HD) and peritoneal dialysis (PD), where blood filtration is the goal. In peritoneal dialysis (PD) the filtration occurs in the abdominal cavity across biological semi-permeable peritoneal membrane, while in hemodialysis (HD) (extracorporeal procedure of blood purification from urea, creatinine and excess water) the transport of water and solutes is achieved in dialysis machine via artificial filters made of polymer membranes.

Despite widespread use of both methods, the reported number of patients with the end-stage renal disease (ESRD) is still growing all over the world (*Roa, et al 2013*). Patients' quality of life is the most important concern for the newer designs of dialysis technique (*Canaud 2013, Ahlmen 2004, Schmaldienst and Hörl 2004, Pereira and Cheung 2000, Vienken 2008, Koda 2011, Progress in Hemodialysis 2011, Fernández 2013, Hoenich 2004*). New developments that can facilitate the dialysis process for patients is a primary goal in the search of new filtering materials (*Vienken 2008,Ronco 2004*) as well as measuring techniques to monitor water redistribution in the organism in the dialysis process – both cost-reducing and non-invasive methods (*Koda 2011, GAMBRO, Fresenius, Asahi Kasei*). The development of feedback systems having the task of automatic control of dialysis and ultrafiltration process to optimize the treatment variables (in particular, the rate of ultrafiltration) requires sensor measurements and computer modeling of the results (*Fernández 2013*).

Synthetic membranes of varying structure and chemical composition are used in contemporary dialyzing devices. A prominent example is the filtration of blood in the artificial kidney systems by GAMBRO (*GAMBRO*), Fresenius (*Fresenius*), Asahi Kasei (*Asahi Kasei*), also see (*Pereira 2000*). Transport of water through dialyzing membranes in parallel with the improved sieving characteristics is a fundamental aim in parallel with the everyday needs and applications in the clinical practice.

Artificial HD membranes are very different in morphology of both skin-layer and core-layer. Modern nanotechnologies permit the fabrication of membranes with given porosity properties and the performance manipulation in the process of membrane preparation (*Karan 2015*). To find the best structure-geometry for the optimization of pore functional shapes, a predictive modeling is needed (*Kanani 2010*). Another help is the 'flows-from-structure'' computerized treatment of micro-tomography images and calculation of permeability value. Such analysis in combination with the experimental data on x-ray tomography has been done in ( *Koivu 2011* ) providing the example of evaluation of flow permeability of porous materials with higher precision where the error of the numerical calculations due to the finite resolution of the tomography images was found not essentially larger than the reported experimental values.

Biofouling is another important factor in design of improved HD filters (*Matsuda 1989, Chanard 2003*). In the latter, proteins - filter surface interaction has a vital role. The example is a heparin coating (*Hinrichs 1997, Shen 2012, 2014*). (Controversial discussion about heparin: double-fold consequences; heparin-related osteoporosis – is the risk decreased? – see, for example, in (*Shen 2014*)). Since the exposure of blood to the surface of the synthetic HD filter leads to the adsorption of proteins (*Chanard 2003*), the interaction of polymers and biomolecules are of key importance for biocompatibility and adhesion to the HD filters. Several aspects at macromolecular transport across the artificial porous membrane should be taken into account: size and geometry (shape) of pores (for example, allow the passage of $\beta_2$ microglobulin (*Hedayat 2012*) and flexibility of the molecules during the channel passage.

Models of permeability and fouling of filters should provide an insight into the mechanism of pore blocking. In this view, the theories of porous materials which model the structure of the materials as random voids interconnected with necks in a 3D network (*Zhdanov 1991*) and include the coverage and blockage of pore entrance (i.e. fouling) in parallel with percolation phenomena (*Schante and Kirkpatric 1971, Blanchard 2000*) could give us useful hints for the analysis.

The conventional HD treatment is an expensive procedure. What is the alternative method for patients with a kidney failure? This is a very reasonable question for developing countries due to the restrictive budget of clinics, limited medical facilities or inaccessibility of hemodialysis equipment.

A new idea on how to remove the toxins during blood filtration has been suggested by Japan International Center for Material Nanoarchitectonics (*Namekawa* 2014). This innovative method of blood purification uses an electrospun network of polymer nanofibers filled with zeolite particles, where zeolites adsorb uremic toxins from the blood due to the microporous structure of the particles. Modeling adsorption in porous media can provide essential theoretical support to the experimental studies. In this respect, one should mention calculations based on empirical equations suggested in (*Saito and Foley* 1991) which include effects of pore curvature studied for the adsorption in microporous zeolites. Another help is the application of lattice density functional methods (*Qajar* 2016) extending the theory from micro- to mesoporous media.

**Summary**

The main goal of the current paper is to provide an integrated view on modeling of molecular transport across dialysis membranes. Hemofiltration is a multiparametric process guided by a complex kinetics of water and solutes penetrating through porous membranes. Current review is focused on selected models, namely, the two –pore- model, the three- pore- model (TPM) and the extended TPM in dialysis applications. Although the peritoneal wall has an extremely sophisticated labyrinthine structure, it was shown (*Öberg Rippe, 2017*) that the TPM model can successfully reflect main features of the peritoneal transport and quantitatively describe the physical mechanism of transmembrane diffusion of water and solutes in agreement with experiments (refs). In particular, it was demonstrated (refs) how the extended TMP model can be used in optimizing automated peritoneal dialysis.

A similar approach also based on nonequilibrium thermodynamics and Kedem-Katchalsky equations has been developed in a number of theoretical works by (*Waniewski; Waniewski et al, Debowska*). There, the kinetic modeling of dialysis originating from one pool- , two pool- and multiple-pool theory of Popovich (Popovich), was generalized to calculate sieving coefficients as well as fractional volume fluid fluxes and albumin time-courses in PD transport (*Debowska* 2011). The exact mathematical solutions obtained within the so-called spatially-distributed model (*Debowska* 2011) have been supplemented with numerical simulations for the tissue in contact with peritoneum. New mathematical tools for controlled dialysis also include a web-based program, a solute calculator program for the quantitative analysis of dialysis measures developed by (*Daurgidas* 2009). Due to the limited text space, these important mathematical models are not included in the current review. The interested readers may find essential formulations in (*Waniewski*; *Waniewski et al; Debowska; Daurgidas etal, Azar*). In addition, modeling of sodium kinetics is of a particular importance (sodium models refs) and should be mentioned.

The effects of tortuosity and shape of porous membrane pathways were not considered in the above-mentioned models. The notions of tortuosity and porosity have been introduced in the theory of hindered diffusion. Specifically, it was shown that the molecular transport across tortuous pores has certain peculiarities for the cake layers formed at fouling of the dialysis filters. The tortuosity and porosity are crucial physical characteristics of hindered diffusion through cake layers (*Armatas et al; Kim and Chen*). The influence of these parameters on tortuous membranes within molecular models are briefly discussed in the second part of the review. A special class of theoretical models which take into consideration the complex structure of dialysis membrane filters belong to stereology (*Chagnac* etal 1999). Mathematical models which account for the effects of pore geometry can be found in (see, for the review, in *Kanani; Ileri et al, 2012*).

The important aim of the current review is to bring together valuable theoretical physics concepts of nonequilibrium thermodynamics and electrical conductivity – diffusion analogy and show how the mathematical models presented here emerge and further develop these fundamental theoretical approaches. Incidentally, general remark should be made here. First, on the analogy between electrical conductivity and diffusion in porous media. This analogy has already been addressed by Lord Rayleigh in his work 'On the influence of obstacles…'(*Rayleigh* 1892) and then, effectively, used by Von Carman (*Carman* 1937) and in a large number of research papers on diffusion through porous materials (for the review, see, for example, a very comprehensive book of *Dullien 1992, other refs, Klinkenberg*, other refs). In addition, there is a deep influence of the theory of dielectric 'mixtures'' originally developed in Maxwell (*Maxwell*), Wagner (*Fricke*), Rayleigh (*Rayleigh*) and Debye (*Debye*) works and later continued by Hanai (*Hanai*), Hanai and Koizumi (*Hanai and Koizumi*), and other researchers (*Fricke and Morse; De La Rue & Tobias*; *Looyenga; Nielsen LE*) on a range of disciplines. In particular, this theory has provided ground for the theory of bioimpedance analysis (BIA) and bioimpedance spectroscopy (BIS), widely used in hemodialysis for the assessment of total body water, as well as intracellular and extracellular water (*De Lorenzo etal 1997; Matthie 2005, 2008*). The latter parameters are crucial for the correct dialysis performance and 'dry weight'' evaluation (refs).

Another comment should be made on the generality and **universality** of above-mentioned fundamental theories. The nonequilibrium thermodynamics framework and the conductivity-diffusion analogy demonstrate general physical principles of mass transport across porous membranes which do not depend on the chemical structure of the system. The kinetic equations for conductivity (or resistivity) and diffusion represent linear relations between the conjugated fluxes and generalized forces while the kinetic coefficients obey the Onsager's principle of symmetry. The notions of 'tortuosity' and 'porosity' (*Dullien* 1992) are also introduced in the theory as universal physical characteristics of the material independent on its chemical composition. These theories lay a foundation for the generalized description of transport of water and solutes in such complex systems as biological and artificial membrane filters used in dialysis. A distinguished example of theoretical soft matter physics' success in soft matter in the theory of polymer dynamics (*DeGennes, Doi &Edwards, Grosberg & Khokhlov*). Within the framework of this approach, physical characteristics such as deformation/rigidity, radius of gyration and persistence length, were introduced for the description of various polymers in solution. The fundamental significance of these notions is their **universality**, i.e. that these parameters do not depend on the specific chemical structure of polymers. Accordingly, the results of theory of polymer dynamics should be useful for the analysis of macromolecular transport across dialysis filter membranes. In particular, the reptation model (*DeGennes, Doi & Edwards*, also, for the review of DNA gel electrophoresis see, for example, excellent paper by Viovy (*Viovy 2000*)) may provide helpful ideas and physical analogies for the quantitative analysis of macromolecular (protein and polysaccharides) diffusion across fouling cake layers.

Theoretical modeling of hindered diffusion is an area of extensive research which cannot be covered in one short review. The basic models which provide valuable insight into physical mechanisms of hindered molecular transport and pore blocking could be found, for example, in (*Ogston; Malone Andersson; Andersson; Hermia; Giddings et al*) also in a very comprehensive review of (*Deen 1987*) as well as in (*Deen et al.*; *Zydney; Bromley et al*; *Ho and Zydney*; *Johnson, Koplik and Dashen*; *Sanandaji et al; Saito and Foley*; *Santos et al; Stawikowska and Livingston*; *Blanchard et al; Bhattacharjee and Sharma*; *Carlsson, Sanandaji et al*). Despite the fact that rigorous theoretical description of water and solute transport in PD and HD membranes remains to be completed, new developments in mathematical modeling comprise studies of diffusion across the asymmetric membrane layers in dialysis filters (*Kim and Chen*). The calculated effective time-dependent diffusivity coefficient (*Hizi and Bergman*) provide useful information about the fast kinetic processes in the system. The search for new improved materials for the membrane filters and best modes of dialysis procedures should take into account the hindered diffusion in cake layers. Notwithstanding unavoidable simplifications of membrane pore's real structure, predictive theoretical modeling can serve as extremely useful mathematical tool in an improved automated dialysis of both types.

**Acknowledgement**

The author is grateful to Carl Öberg for the stimulating and fruitful discussions on current publication, as well as to Leonid Gorelik for his support and help in manuscript preparation. Special thanks to Tatiana Shilnova and Mikhail Shilnov for the help in manuscript preparation.